\documentclass{aa}  

\usepackage{graphicx}
\usepackage{txfonts}
\usepackage{booktabs}
\usepackage{amsmath}
\usepackage{natbib}
\usepackage{float}
\usepackage{subcaption}
\usepackage[symbol]{footmisc}

\usepackage[usenames,dvipsnames,svgnames,table]{xcolor}
\usepackage[
breaklinks=true,
hyperindex=true,
colorlinks=true,
linkcolor=Sepia,
citecolor=Purple]{hyperref}

\makeatletter
\renewcommand*\aa@pageof{, page \thepage{} of \pageref*{LastPage}}
\makeatother

\newcommand{\uSFR}{M$_\odot\,$yr$^{-1}$}

\newcommand{\obacht}[1]{}

\newcommand{\changeSummary}[1]{\obacht{#1}}
\newcommand{\same}{\obacht{This part is the same as in the first submission \\~\\}}

\begin{document} 

   \title{Introducing the Rhea\thanks{ 
   Rhea (P$\epsilon\alpha$) is a goddess and Titanide from Greek mythology. Her milk formed the Milky Way, as described by \citet{Rhea1, Rhea2}. To prevent her husband Cronus from devouring her newborn son Zeus, she hands him a swaddled stone. 
   When he asks her to nurse the child a last time, she presses her breast and the flowing milk forms the Milky Way.\\} simulations of Milky-Way-like galaxies}
   \subtitle{I: Effect of gravitational potential on morphology and star formation}

   \author{Junia Göller \inst{1}\thanks{junia.goeller@uni-heidelberg.de}
   \and
   Philipp Girichidis \inst{1}\thanks{philipp@girichidis.com}
   \and
   Noé Brucy \inst{1,2}\thanks{noe.brucy@ens-lyon.fr}
   \and
   Glen Hunter \inst{1}
   \and
   Karin Kjellgren \inst{1}
   \and
   Robin Tress \inst{2}
   \and
   Ralf S.\ Klessen \inst{1,4,5,6}
   \and
   Simon C.~O.~Glover \inst{1}
   \and 
   Patrick Hennebelle \inst{7}
   \and
   Sergio Molinari \inst{8}
   \and
   Rowan Smith \inst{9}
   \and 
   Juan D.~Soler \inst{8}
   \and
   Mattia C.~Sormani \inst{10}
   \and
   Leonardo Testi \inst{11, 12}
   }

   \institute{
    Universit\"{a}t Heidelberg, Zentrum f\"{u}r Astronomie, Institut f\"{u}r Theoretische Astrophysik, Albert-Ueberle-Str.\ 2, 69120 Heidelberg, Germany
    \and
    Centre de Recherche Astrophysique de Lyon UMR5574, ENS de Lyon, Univ. Lyon1, CNRS, Université de Lyon, 69007 Lyon, France
    \and
    École Polytechnique Fédérale de Lausanne, Observatoire de Sauverny, Chemin Pegasi 51, CH-1290 Versoix, Switzerland
    \and 
    Universit\"{a}t Heidelberg, Interdisziplin\"{a}res Zentrum f\"{u}r Wissenschaftliches Rechnen, Im Neuenheimer Feld 225, 69120 Heidelberg, Germany
    \and
    Harvard-Smithsonian Center for Astrophysics, 60 Garden Street, Cambridge, MA 02138, U.S.A. \label{CfA}
    \and
    Elizabeth S. and Richard M. Cashin Fellow at the Radcliffe Institute for Advanced Studies at Harvard University, 10 Garden Street, Cambridge, MA 02138, U.S.A. \label{Radcliffe}
    \and
    Université Paris-Saclay, Université Paris Cité, CEA, CNRS, AIM, 91191, Gif-sur-Yvette, France
    \and
    Istituto di Astrofisica e Planetologia Spaziali (IAPS), INAF, Via Fosso del Cavaliere 100, 00133 Roma, Italy
        \and
    SUPA, School of Physics and Astronomy, University of St Andrews, North Haugh, St Andrews, KY16 9SS, UK
    \and
    Como Lake centre for Astrophysics (CLAP), DiSAT, Universit\'{a} degli Studi dell'Insubria, Dipartimento di Scienza e Alta Tecnologia, Via Valleggio 11, 22100 Como, Italy
    \and
    Alma Mater Studiorum Università di Bologna, Dipartimento di Fisica e Astronomia (DIFA), Via Gobetti 93/2, 40129 Bologna, Italy 
    \and
    INAF – Osservatorio Astrofisico di Arcetri, Largo E. Fermi 5, 50125 Firenze, Italy
    }

   \date{}

\abstract
{The Milky Way is a complex ecosystem, for which we can obtain detailed observations probing the physical mechanisms determining the interstellar medium. For a detailed comparison with observations, and to provide theories for missing observables, we need to model the Milky Way as closely as possible. However, details of the Galactic structure are not fully defined by observations, raising the need for more generalized models.
With the Rhea simulations we present a set of Milky Way like simulations, containing detailed physics of the interstellar medium, as well as star formation and stellar feedback. We conduct two simulations that differ in the gravitational potential: one fitted to several structural details derived from observations, the other just reproducing the most basic quantities. 
We find little difference in the overall morphology except for the bar region, which funnels gas towards the Galactic inner region and therefore prevents quenching in the center. Despite differences with galacto-centric radius, the global star formation rate is almost identical in both setups.
A spiral arm potential does not influence properties of groups of formed stars. A bar potential, however, lowers size and formation time of those associations.
We therefore conclude for a spiral arm potential to have little influence on star formation in the Galaxy, except for producing long-lived spiral structures instead of transient ones. A Galactic bar potential has noticeable influence on star formation mainly within the innermost 2.5~kpc.}


\maketitle
\nolinenumbers
\section{Introduction} \label{sec:intro}
\changeSummary{Introduction completely rewritten with a focus on the Galactic potential and star formation.}

The Milky Way, as a typical spiral galaxy of the type that dominates the star formation in our present-day universe \cite[][]{Kennicutt2012, Bland-Hawthorn2016}, is a unique laboratory for physical processes of galaxy evolution. It provides the opportunity for direct observation of processes such as the formation of molecular clouds, stars, and stellar feedback. In no other galaxy can we study these processes in as much detail as in our own. 

However, the galactic context of star formation and corresponding stellar feedback is still poorly understood, even within the Milky Way. Especially the influence of the galactic potential and galactic dynamics on local star formation is an open debate. Indeed, star formation is known to be a very inefficient process \citep{Girichidis2020}. A key question is whether this inefficiency is entirely due to the influence of stellar feedback, or whether galactic dynamics and the galactic potential also play an important role in regulating the star formation efficiency, as star formation does not follow an universal rule.

The central molecular zone (CMZ) in the center of the Milky Way, for example, is a compact region of about $3-7\times10^7$~M$_\odot$ \citep{Molinari2011, Tokuyama2019} of cold, dense clouds, fueled by the dust lanes caused by the Galactic bar potential. The SFR of the CMZ, however, is observed to be just $0.06-0.14$~M$_\odot$~yr$^{-1}$ \citep{YusefZadeh2009, Immer2012, Longmore2013}, more than a factor of 10 smaller than  what would be predicted from its gas mass and density \citep{Longmore2013, Kruijssen2014}. Simulations \citep[e.g.][]{SeoKim2013, SeoKim2019, Robin2, Moon2021} suggest that this is because the SFR is more closely correlated to the gas inflow rate (which is largely controlled by the bar potential) than to the mass of the region.

Another example are spiral arms, whose role in star formation is still heavily debated. It is unclear if they mainly serve as triggers of star formation by increasing the density of incoming gas and the star formation efficiency (SFE), or if they simply organize star-forming gas with no impact on the star formation process itself. As an example, \citet{Foyle2010} found very similar SFEs in arm and interarm regions of NGC 5194 and NGC 628, whereas \citet{Seigar2002} found significant enhancement of SFR in the vicinity of spiral arms, arguing for triggered star formation. Numerical simulations \citep{Dobbs2011,KimKim2020} tend to find no triggering of star formation in spiral arms. The question is even harder to answer without the knowledge of whether the spiral structures are long-lived \citep[as suggested by the density wave theory,][]{LinShu1964, Shu2016} or whether they are transient features \citep[as suggested by the self-propagating star formation model,][]{Mueller1976, Gerola1978}. 

For the Milky Way itself, it is hard to estimate the true Galactic potential and also star formation properties anywhere outside the greater solar neighborhood, because of our particular observational placement in it. The Galaxy has been sampled across a wide range of wavelengths, revealing many facets of its complex system. Different surveys target dust (e.g. ATLASGAL, \citealt{Atlasgal} or Hi-GAL, \citealt{Hi-GAL}), the molecular gas (e.g.\ as sampled by emission from carbon monoxide: SEDIGISM, \citealt{Sedigism}, CHIMPS, \citealt{Chimps}, FQS, \citealt{Fqs}), the atomic phase (e.g.\ as traced by H{\sc i} emission: THOR, \citealt{Thor}, and VGPS, \citealt{Vgps}), or the ionized component of the ISM (e.g.\ as seen by  eROSITA, \citealt{Predehl2021}, or twenty years earlier by ROSAT, \citealt{Voges1999} for the hot ionized medium, or in the WHAM sky survey \citep{Haffner2003} for the warm ionized medium). For stars see \citet{Gaia-DR3} (or \citealt{GLIMPSE, MIPSGAL} for infrared surveys), and for nearby star-forming regions e.g.\ \citet{Glostar}. However, basic Galactic properties are still under debate, like how many spiral arms are present \citep{HouHan2014, Drimmel2000, Reid2019} and the exact structure and relative position of the Galactic bar \citep[see e.g.][]{Nishiyama2005, Cabrera2008, Vanhollebeke2009, Majaess2010}. One therefore might wish for a more general approach in terms of potential and structure when modeling a Milky Way like galaxy.

Observations are complemented by numerous simulations, aiming to explain and supplement observations. In cosmological contexts, Milky Way like galaxies have been studied in Illustris \citep{MWIllustris}, IllustrisTNG \citep{MWTNG}, EAGLE \citep{MWEagle}, FIREbox \citep{MWFirebox} and Simba \citep{MWSimba}. With this, the full cosmological formation process of a galaxy can be followed. However, they are usually too coarse to study individual star-forming sites in detail. In `zoom-in` simulations (e.g. Auriga \citep{Auriga}, VINTERGATAN \citep{Vintergatan}, APOSTLE \citep{Apostle}, NIHAO \citep{Nihao}, Latte \citep{Latte}), where promising structures from large-scale simulations are re-simulated at higher resolution, the formation of smaller-scale structures can be studied, as well as components like the Galactic magnetic field. These simulations have expanded our understanding of processes shaping galaxies such as the Milky Way. However, they usually still lack the resolution to resolve small-scale processes that can only be observed in the local Galaxy, such as the formation of molecular clouds and cores.

A third approach is to simulate the Milky Way as an isolated galaxy, such that much higher resolutions can be reached. This approach was used for example by \citet{Pettitt2014}, who use the SPH code \textsc{Phantom} to simulate a Milky Way-like galaxy with a modular potential of bulge, halos, disk and spiral arms in a 13~kpc disk, and recreate morphological features of the Milky Way in CO emission. They take into account ISM cooling and a simplified hydrogen and CO chemistry, but no star formation and stellar feedback and no magnetic field. With the AMR  code \textsc{Ramses}, \citet{Renaud2013} simulate a 28~kpc gaseous disk of a Milky Way analogue with star formation and stellar feedback for a few cloud lifetimes. They adopt a dynamic potential including the dark matter (DM) halo, spheroid and bulge, thin and thick disk. Non-axisymmetric features such as the bar and spiral arms are formed during the run from instabilities in the velocity profiles. The simulations of \citet{Jeffreson2020} use the moving-mesh code \textsc{Arepo} to study giant molecular clouds in the disk of a Milky Way-like galaxy. They test the impact of several potentials (consisting of different compositions of the stellar bulge, stellar disk and DM halo) on a gaseous disk with an exponential density profile (scale radius 7.4-7.7~kpc). Star formation, stellar feedback, and non-equilibrium chemistry of hydrogen, carbon, and oxygen are considered. However, they ignore the Galactic bar in their potential. \citet{Wibking2023} simulate a Milky Way like Galaxy with \textsc{Gizmo} to study the Galactic magnetic field. The include full magnetohydrodynamical (MHD) treatment, a chemical network, star formation, photoionization and supernova feedback. However, their simulation, as well, lacks a Galactic bar. The same applies for \citet{Konstantinou2024}, a study of the $B-\rho$ relation in a Milky Way like galaxy, using adaptive mesh refinement with MHD, chemistry, star formation and supernova feedback. They include a DM halo, thin stellar disk (truncated at 12~kpc), gaseous disk (truncated at 15~kpc) and gaseous halo.

As a complementary step, \citet{Robin1} and \citet{Robin2} (again using \textsc{Arepo}) only simulate the inner disk and the central molecular zone (CMZ) of the Milky Way with sub-pc resolution. In their simulation of the innermost 5~kpc of the Galactic disk, they allow for star formation and feedback and include a chemical network following the different states of hydrogen, oxygen, ionized carbon and CO. Another way to overcome the cost of simulating a full galaxy is to simulate only a small region of it. This allows higher resolution, broader parameter study and more accurate but more expensive treatment of the physics of the ISM, such as radiative transfer \citep[eg.][]{walchSILCCSImulatingLifeCycle2015,girichidisSILCCSImulatingLifeCycle2016,kimThreephaseInterstellarMedium2017,collingImpactGalacticShear2018,KimKim2020,brucyLargescaleTurbulentDriving2020,rathjenSILCCVIIGas2023}. This technique has the drawback that it neglects the connection with the larger galactic scales, which probably have an important influence on the SFR \citep[e.g.][]{brucyLargescaleTurbulentDriving2020} and the structure of the ISM \citep[e.g.][]{colmanSignatureLargescaleTurbulence2022}.
A compromise is to realise zoom-in simulations from isolated galaxy simulations, which consist in having high-resolution only in a selected comoving region of the disk \citep{smithCloudFactoryGenerating2020,fenschUniversalGravitydrivenIsothermal2023,zhaoFilamentaryHierarchiesSuperbubbles2024}.

However, a simulation of the whole Milky Way galaxy that allows for a study of the influence of the Galactic potential on star formation is still missing, as either essential parts of the potential are not included, or the simulation time is too short to produce meaningful statistics. Here we present the first set of simulations from the Rhea simulation suite, targeted to fill this gap. Our simulations include an elaborate external Milky Way potential (with non-axisymmetric features of a bar and spiral arms) that allows for a close match to observations of dynamical properties, as well as star formation and supernova feedback. To allow us to investigate the influence of the potential, we model the same set of initial conditions also with a potential that just reproduces the most basic features of the velocity curve.
This paper is the first in a series that introduces the main setup and discusses the numerical techniques. Here, we focus on hydrodynamical simulations, whereas follow-up studies include magnetic fields and cosmic rays. As a first application, we compare the influence of the potential on the basic properties of the galaxy, i.e.\ its general morphology and gas distribution and especially its star formation and stellar feedback.

The paper is organized into the following sections: In Section \ref{sec:methods}, we begin with the general simulation methods and set up, the adopted star formation and stellar feedback routines, the different potentials and further simulation details. We describes the morphological and structural properties in Section \ref{sec:results:morphology}. In Section \ref{sec:results:sf} we analyze the differences in star formation arising between the two studied potentials. We briefly discuss the caveats of our study in Section \ref{sec:caveats} and in Section \ref{sec:conclusions} we summarize our findings.

\section{Methods}\label{sec:methods}

\changeSummary{{Little differences, mainly from removing MHD simulations.}}

We model an isolated Milky-Way like galaxy with a fixed external potential. In the following, we present our numerical setup as well as the parameters that we use.

\subsection{Numerical framework}\label{sec:numerics}
\changeSummary{{References to MHD removed.}}

In this work, we present a set of 2 simulations of a Milky-Way-like galaxy (+2 simulations for a resolution study, specified in Table \ref{tab:simulations}). We use the moving-mesh code \textsc{Arepo} \citep{arepo, arepo_public}, which solves the equations of hydrodynamics while also accounting for the gravitational accelerations produced by the gas and by collisionless components such as stars and dark matter. The fundamental equations of hydrodynamics 
in a non-cosmological environment ($a=1$) can be written as \citep{arepo_public, Springel2013}:
\begin{align}
\frac{\partial \rho}{\partial t}+\nabla \cdot \left(\rho \boldsymbol{v}\right) &= 0,\\
\frac{\partial \rho \boldsymbol{v}}{\partial t} + \nabla \cdot \left(\rho \boldsymbol{v} \boldsymbol{v}^\mathrm{T}+\boldsymbol{I}p_\mathrm{tot}\right)&=-\rho \nabla \Phi,\\
\frac{\partial E}{\partial t}+\nabla \cdot \left(\boldsymbol{v}\left(E+p_\mathrm{tot}\right)\right)&=
-\rho \left(\boldsymbol{v}\cdot \nabla \Phi\right)+\mathcal{H}-\Lambda,
\end{align}
with the total energy (per unit volume) and pressure being
\begin{align}
    E&=\rho e_\mathrm{th}+\frac{1}{2}\rho \boldsymbol{v}^2,\\
    p_\mathrm{tot}&=\left(\gamma-1\right)\rho e_\mathrm{th}.
\end{align}
Here, $\rho$ is the mass density, $\boldsymbol{v}$ is the flow velocity vector, $\mathcal{H}$ and $\Lambda$ are the radiative heating and cooling terms respectively, and  $e_\mathrm{th}$ is the thermal energy per unit mass. The gravitational potential $\Phi$ is the sum of the external potential, $\Phi_\mathrm{ext}$, described below, the gas self-gravity, $\Phi_\mathrm{gas}$, and the potential due to the star particles, $\Phi_\mathrm{stars}$. We set the adiabatic index to $\gamma =5/3$ for all gas, even if it is molecular. This is justified because the vast majority of the molecular gas in our simulations has a temperature $T < 200$~K, which is too low to excite the internal degrees of freedom of the H$_{2}$ molecule. 

The hydrodynamical equations are solved on a 3D time dependent Voronoi mesh. The mesh generating points defining the Voronoi tessellation move at the local fluid velocity, such that the grid can follow the fluid. The cell mass is held approximately constant, and therefore the cell volume continuously adapts to regions of differing density, making \textsc{Arepo} a quasi-Lagrangian scheme.

\begin{table}[t]
    \centering
    \begin{tabular}{lccc}
    \toprule
     & mass resolution 
     & potential  \\
     & [M$_\odot$] &  \\
     \midrule
     F3000HD & 3000 
     & flat  \\
     MW3000HD & 3000 
     & MW  \\
     F1000HD & 1000 
     & flat \\
     MW1000HD & 1000 
     & MW \\
     \bottomrule
    \end{tabular}
    \caption{Specifications of the simulations presented in this work.}
    \label{tab:simulations}
\end{table}

\subsection{Chemistry and thermodynamics}\label{sec:methods:chemistry}
\changeSummary{{Section refined to clarify the purpose of the chemical network.}}

We use a chemical network to model the non-equilibrium chemical composition of the gas. At our fiducial resolution (see Section \ref{sec:methods:resolution} below), we do not expect to be able to accurately model the atomic-to-molecular transition (see e.g.\ \citealt{seifried17} or \citealt{joshi19} for a discussion of the required resolution) and hence any results regarding the H$_{2}$ or CO content of the gas must be treated with great caution. Our decision to follow the chemical evolution of the gas despite this fact is motivated by two main considerations. First, and most importantly, modelling the chemistry allows us to track the fractional ionization of the gas. This plays an important role in determining the efficiency of both photoelectric heating and C$^{+}$ cooling \citep{wolfire95}, and tracking its value on the fly in the simulation therefore allows us to model the thermal evolution of the gas much more accurately than if we were using a simple tabulated cooling function. Second, we plan to follow up on our current calculations in future work in which we will perform much higher resolution `zoom-ins' of selected regions. Having chemical information already available in the output from our current simulations will therefore greatly facilitate their use as initial conditions for these upcoming calculations.

In our simulations, we use the NL97 chemical network of \citet{Glover2012}. This combines the hydrogen chemistry network from \citet{Glover2007a, Glover2007b} and a highly simplified CO chemistry network introduced by \citet{NL97}. To model the H$_2$ and CO self-shielding and the dust shielding from the UV interstellar radiation field, we use the \textsc{TreeCol} algorithm developed by \citet{Treecol}. We consider a spatially and temporally constant background radiation field, using solar neighbourhood values for the strength and spectral shape from \citet{Draine78} in the UV and \citet{mathis1983} in the optical and infrared. We also take into account cosmic ray ionization, with a rate $\zeta_{\rm H}$\,$=$\,$3\times 10^{-17} \: {\rm s^{-1}}$ for atomic H, and suitably scaled values for other chemical species.

Apart from adiabatic expansion and contraction, which are treated in the standard hydrodynamics of \textsc{Arepo} \citep{arepo}, we take into account several additional radiative and chemical heating and cooling processes \citep{Clark2019}. The main heating processes include the photoelectric effect, H$_2$ photodissociation, chemical heating due to UV pumping of H$_{2}$ and H$_{2}$ formation on dust grains, and heating associated with cosmic ray ionization. The main cooling processes include fine structure line emission from C$^{+}$ and O, 
rovibrational line emission from H$_{2}$ and CO,  collisional ionization of atomic H and dissociation of hydrogen, gas-grain energy transfer, ion recombination on grain surfaces, and Compton cooling. A full list of all included processes is given in \citet{Glover2010}, with later additions and modifications described in \citet{Glover2012} and \citet{Mackey2019}.

Throughout the simulation, we impose a temperature floor.
During phase I (see Sec. \ref{sec:methods:preproc}), we set this temperature floor to 100\,K, to prevent the gas from cooling and forming a large number of star particles in the same simulation region. In phase II, we drop it to 20~K. This is lower than the characteristic temperature of the gas at the highest densities reached in our simulations (see Fig.\ \ref{fig:phaseplot}), and so we do not expect this temperature floor to significantly affect the dynamical behaviour of the gas.

\subsection{Star formation}\label{sec:methods:sf}
\changeSummary{Correction of a typing error in the second to last paragraph.}

To represent clusters of stars, we use star particles, i.e.\ point-like structures in the code, that form out of gas and thereafter interact only via gravitational forces. At the resolution used in our simulations, each star particle has a mass much greater than a typical star and, therefore, represents multiple stars. With the information about the star particle, we also store a list of the masses and lifetimes of the massive ($M > 8 \: {\rm M_{\odot}}$) stars that each particle represents (see Section \ref{sec:methods:sfb}). This is used to determine the stellar feedback produced by each star particle, as explained in Section~\ref{sec:methods:sfb}. In contrast to the sink particles used in some simulations of galactic-scale star formation \citep[e.g.][]{tress2020simulations}, which represent a mix of dense gas and stars, the star particles used here represent only stars, i.e.\ the full mass of a gas cell gets converted into stellar mass, no hidden gas remains in the star particle.

To decide whether gas within an active grid cell forms a star particle during a given timestep, we investigate the gas in terms of its Jeans instability. This approach differs from a density-threshold approach \citep[such as adopted in e.g.][]{Renaud2013, Jeffreson2020} and therefore prevents artificial star formation in hot, compressed gas as it might occur in feedback regions. We first calculate the Jeans mass $M_J$ of the cell via 
    \begin{align}
        w&=\frac{5e_\mathrm{th}(\gamma -1)}{G},\\
        v&=\frac{3}{4\pi \rho},\\
        M_J&=w^{3/2}v^{1/2},
    \end{align}
where $e_\mathrm{th}$ is the specific internal energy of the cell and $G$ is the gravitational constant. If the cell's mass exceeds 12.5\% of its local Jeans mass, we flag the cell as possibly star-forming. This number is chosen such that star formation starts when the Jeans mass is resolved by less than 8 cells. If the mass exceeds the Jeans mass, we immediately force it to form a star particle. By doing so, we prevent dense gas from accumulating on the grid and forcing small time steps.
This procedure was first introduced in \citet{smith2021efficient}.

If the cell was flagged as possibly star-forming,  we calculate its free-fall time $t_{\rm ff}$ and star formation rate (SFR)
\begin{align}
        t_{\rm ff}=\sqrt{\frac{3\pi}{32 G \rho}},\\
        \mathrm{SFR}=\epsilon \frac{M_\mathrm{cell}}{t_{\rm ff}},
    \end{align}
where $M_\mathrm{cell}$ is the mass of gas in the cell. 
We set the star formation efficiency per free-fall time, $\epsilon$, to a value of 1\%, motivated by recent observational determinations \citep[see e.g.][]{Sun2023}. However, we have varied this parameter between $0.1$\% and $10$\% and found it to have no strong influence on the overall SFR in the simulation, owing to efficient self-regulation of star formation. 

Given the SFR of the cell, we then determine its probability $p$ of forming a star during the current timestep using the same expression as in \citet{springel2003cosmological}:
\begin{align}
        \lambda&=\mathrm{SFR}\frac{\mathrm{\Delta t}}{M_{\rm cell}},\\
        p&=\frac{M_\mathrm{\rm cell}}{M_\mathrm{starP}}(1-\exp(-\lambda)),
    \end{align}
where $\Delta t$ is the timestep of the cell. $M_\mathrm{starP}$ is the desired mass of the star particle, which equals $M_\mathrm{cell}$ in most cases. That means that usually a full gas cell is converted into a star particle. However, if the mass of the cell exceeds the target mass (see Sec. \ref{sec:methods:resolution}) by more than a factor of 2, i.e.\ if the cell would have been split into two in the next timestep, only half of the cell's mass gets converted into a star particle, and the other half remains as gas.

The timestep of star-forming cells is limited to 
\begin{align}
    \Delta t<0.1\frac{M_\mathrm{cell}}{\mathrm{SFR}}.
\end{align}
With that, $p\lesssim0.1 M_\mathrm{cell} M_\mathrm{starP}^{-1}$, and therefore the probability for a star particle to form cannot exceed 1 until the target mass chosen for the star particle is ten times smaller that the mass of the gas cell. Since we set the target mass for the star particles to equal the target mass for the gas cells, and the code prevents gas cells from exceeding their target mass by more than a factor of 2, that does not happen.

After the star formation probability is calculated, we draw a random number, $x_\mathrm{SF}$, from a uniform distribution between 0 and 1, and in case $x_\mathrm{SF}<p$, the full gas cell gets converted into a star particle, i.e.\ all of the gas in the cell is converted into stars. Otherwise, it remains as gas.

\subsection{Stellar feedback}\label{sec:methods:sfb}
\same
In these simulations, we use the supernova feedback routine described in \citet{tress2020simulations}, with some adjustments toward the used star particles.

In the first step, the star particles are populated with individual stars by sampling from the initial mass function (IMF) following the algorithm described in \citet{sormani2017simple}. For the IMF-sampling we use the high-mass end of the Kroupa IMF \citep{Kroupa} with an efficiency of $\epsilon=1$, i.e.\ the whole mass of the star particle is in stars with no locked-up gas. Stars with a mass between $8\,\mathrm{M}_\odot$ and $120\,\mathrm{M}_\odot$, which are thought to explode as supernovae (SN), get assigned a lifetime in accordance with Table\,25.6 in \citet{maeder2008physics}; we linearly interpolate between the different mass bins in the table. After this lifetime the star explodes as a SN at the location of its star particle.

The SNe explode in one of two modes: energy or momentum injection. The decisive factor here is the radius of the supernova remnant at the end of the Sedov-Taylor phase. At solar metallicity, this is \citep{Blodin1998, tress2020simulations}
\begin{align}
    R_\mathrm{ST}=19.1\left(\frac{\bar{n}}{1 \: \mathrm{cm}^{-3}}\right)^{-7/17}\,\mathrm{pc},
\end{align}
where $\bar{n}$ is the local mean number density. If $R_\mathrm{ST}>R_\mathrm{inject}$, an energy of $10^{51}\,$erg is injected isotropically in the injection region (cells within $R_\mathrm{inject}$) as thermal energy and the contained gas is fully ionized. Otherwise the terminal momentum \citep[derived e.g.\ in][]{Gatto2015, KimOstriker2015, Martizzi2015}
\begin{align}    
p_{\rm fin}=2.6\times10^5n^{-2/17}\,\mathrm{M}_\odot\,\mathrm{km}\,\mathrm{s}^{-1}
\end{align}
is injected directly into the cells within the injection region. In this case, the temperature or ionization state of the gas is not changed. We use an injection radius of $R_\mathrm{inject}=100\,\mathrm{pc}$. This radius ensures a sufficient number of cells within the injection radius.

For SNe in very low-density environments, $R_\mathrm{ST}$ is large, and consequently, the injected thermal energy can lead to unphysically high temperatures. In this case we limit the energy injection as follows: If the temperature of a SN-heated cell, $T_\mathrm{current}=e_\mathrm{th}(\gamma-1)\mu m_p/k_B$, is larger than $10^7$~K, we do not inject any more energy. If the estimated temperature after energy injection
\begin{align}
    T_\mathrm{estimated}=\frac{(e_\mathrm{th}+e_\mathrm{injected})(\gamma-1)\,\mu m_p}{k_\mathrm{B}},
\end{align}
is larger than the threshold value, we reduce the injected energy, $e_\mathrm{injected}$, by a factor of
\begin{align}
   f_l=\frac{(T_\mathrm{cut}-T_\mathrm{estimated})k_\mathrm{B}}{(\gamma-1)\,\mu m_p e_\mathrm{injected}}+1.
\end{align}

\subsection{Gravitational Potentials}
\label{sec:methods:potential}
\changeSummary{We added a description of self-gravity in the simulations for clarity.}

For our setups the gravitational potential is the sum of an external potential $\Phi_\mathrm{ext}$ and the contribution due to self-gravity of the gas and stars $\Phi_\mathrm{sg}$,
\begin{equation}
    \Phi_\mathrm{tot} = \Phi_\mathrm{ext} + \Phi_\mathrm{sg},
\end{equation}
where the latter is found by solving the Poisson equation
\begin{equation}
    \Delta \Phi_\mathrm{sg} = 4\pi G (\rho_\mathrm{gas} + \rho_\mathrm{stars})
\end{equation}
using the standard gravitational tree in \textsc{Arepo} \citep{arepo_public}.

We use parameterized external potentials for the components that we do not include such as dark matter and the low mass stars as well as old populations of stars, because this allows for a better tuning towards Milky Way features. This is different from a live potential, as used by e.g. \citet{Duran-Camacho2024}. We compare two different setups concerning the external potentials: a logarithmic potential, resulting in a flat velocity curve \citep{binney2008galactic}, and a realistic potential fine-tuned to match the known properties of the distribution of dark matter, gas and stars in the Milky Way \citep[MW,][]{Glen}. We will refer to the potentials as `flat' and `MW' from here onwards. 

\subsubsection{Flat external potential}
The `flat' external potential is a logarithmic potential from \citet{binney2008galactic} (Eq. 2.71a), written as 
\begin{equation}
    \Phi_L=\frac{1}{2}v_0^2\ln\left(R_c^2+R_\mathrm{gal}^2+\frac{z^2}{q_{\Phi}^2}\right)+\mathrm{const.},
\end{equation}
where $R_c=100\,$pc and $v_0=220\,\mathrm{km}/\mathrm{s}$ are constants and $q_{\Phi}=0.8$ is the axis ratio of the equipotential surfaces. This results in a circular velocity at the Galactocentric radius $R_{\rm gal}$ in the equatorial plane (eq. 2.71b of \citealt{binney2008galactic}) of 
\begin{equation}
    v_c=\frac{v_0R_\mathrm{gal}}{\sqrt{R_c^2+R_\mathrm{gal}^2}}.
\end{equation}

\subsubsection{MW external potential}
\label{subsec:mw_ext_pot}
The MW external potential of \citet{Glen} is generated from analytic density profiles, including 
\begin{itemize}
    \item the central black hole (SgrA*)
    \item the nuclear stellar cluster
    \item the nuclear stellar disk
    \item the Galactic bar
    \item axisymmetric and non-axisymmetric (spirals) disk components
    \item the DM halo
\end{itemize}
using the \textsc{Agama} code \citep{agama}. The chosen model enforces 4 spiral arms with a pattern speed of $\Omega_\mathrm{spa}=-22.5$~km~s$^{-1}$~kpc$^{-1}$ and a Galactic bar with a pattern speed of $\Omega_\mathrm{bar}=-37.5$~km~s$^{-1}$~kpc$^{-1}$ and is fully described in \citet{Glen}. Figure~\ref{fig:potentials} depicts the non-axisymmetric components of the external potential, the spiral arm (left) and the bar (right) at 2500~Myr. We introduce these non-axisymmetric components linearly within the first 150 Myr of the simulation to avoid transients.
The model presented in \citet{Glen} includes the influence of the gas disk. However, in our simulations we exclude this component of the potential and replace it with the self-consistently computed one of the gaseous disk in our simulation box.

\subsection{Resolution}\label{sec:methods:resolution}
\changeSummary{{Fig. 1 was moved to the Appendix and all references to the figure were refined.}}

\textsc{Arepo} in its standard setting uses a mass resolution, not a volume resolution for the gas cells. Cells with masses that surpass a user-set target mass by more than a factor of $2$ get split into smaller cells, while those that have a mass lower than the target mass by a factor of $0.5$ get removed. We run our simulations with a fiducial target mass of  $3000\,\mathrm{M}_\odot$, but also perform two
comparison runs with a lower target mass of $1000\,\mathrm{M}_\odot$ (see Table \ref{tab:simulations}).

In addition, we also limit the maximum and minimum volumes that our cells are allowed to have, refining or de-refining as appropriate to enforce these. Within the whole box, the minimum cell volume is set to $1\,\mathrm{pc}^3$ and the maximum to $2\,\mathrm{kpc}^3$.  
It is refined if a cell exceeds this volume by a factor of 2.
Moreover, in the area of the Galactic disk, i.e.\ in a cylinder with $30\,\mathrm{kpc}$ in radius and $2\,\mathrm{kpc}$ in thickness, located at the center of the box, we set an upper volume limit of $10^6\,\mathrm{pc}^3$. This ensures a sufficient volume resolution within the Galactic disk. 
In Fig.\ \ref{fig:density-volume_with_highres}, we show the density-volume distribution of gas cells at about 2500~Myr for F3000HD, as a representative example. The mass refinement and the total box volume refinement are strictly adhered to, however, one can see some disk cells 
being larger than allowed by their refinement criterion. 
We explain this by cells existing near the boundary of refinement, moving into the refinement region. Those cells may need multiple refinement cycles to meet the refinement criterion, depending on their size, as during each refinement cycle they can only refine once.

The gravity solver of \textsc{Arepo} treats each gas cell as a point mass with an adaptive softening length, chosen from tabulated softening lengths depending on the cell size. In our simulations, the tabulation starts at $0.01$~pc and has a logarithmic spacing with a multiplicative factor of $1.15$. The softening length is then chosen as the closest value from this tabulation to $2\times r_\mathrm{cell}$, where $r_\mathrm{cell}$ is the radius of a sphere equivalent in volume to the gas cell. Star particles are also treated as point masses, with a fixed softening length of $6.47$~pc.

\subsection{Initial conditions}\label{sec:methods:ics}
\changeSummary{{Minimal adjustments.}}

We start our simulations with a smooth gaseous disk with a density distribution given by 
\begin{align}
    \rho_\mathrm{gal} \left(R_\mathrm{gal}, z\right)=\frac{\Sigma_0}{4z_{\rm d}}\mathrm{exp}\left(-\frac{R_{\rm m}}{R_\mathrm{gal}}-\frac{R_\mathrm{gal}}{R_{\rm d}}\right)\mathrm{sech}^2\left(\frac{z}{2z_{\rm d}} \right)
    \label{equ:density_profile}
\end{align}
in cylindrical coordinates \citep{sormani2019geometry}, with $z_{\rm d}=85\,$pc, $R_{\rm d}=7\,$kpc, $R_{\rm m}=1.5\,$kpc and $\Sigma_0=50\,\mathrm{M}_\odot\,\mathrm{pc}^{-2}$. The total mass of the gas disk is $\sim10^{10}$~M$_\odot$.
We truncate this distribution at a radius of 30~kpc, as expected from the maximum radial extent of HI in the Milky Way \cite[see, for example,][]{Kalberla2008,AtomicHydrogenInTheMW}.
Further out, the density is set to a minimum density $\rho_{\rm min} = 10^{-31}\,\mathrm{g}\,\mathrm{cm}^{-3}$. We do not impose a cut in the $z$-direction, but set a minimum density of $10^{-31}\,\mathrm{g}/\,\mathrm{cm}^{-3}$ here as well. The gas is set to solar metallicity and a gas-to-dust ratio similar to the local ISM. We use an initial abundance ($A\left(X\right)=\log\left(X/H\right)+12$) of $8.15$ for carbon and $8.5$ for oxygen \citep{Simbach2000}.
The disk is embedded in a box of 150~kpc side length with periodic boundary conditions. The large box size prevents outflows from reaching the box boundaries and, therefore, boundary effects from happening. The gas initially has a temperature of $1.3\times10^4$~K throughout the whole box. After generating the simulation box with this density distribution, we impose the velocity corresponding to the used potential on the gas, while accounting for the pressure gradient.

\subsection{Simulation phases}\label{sec:methods:preproc}
\changeSummary{{Minimal adjustments.}}

Without turbulence, our initial disk with its smooth density distribution would collapse immediately, forming a huge number of stars, which, with their feedback, would destroy the Galactic disk. To prevent that, we initially set a temperature floor of 100~K to prevent the gas from cooling too much and induce turbulence in an initial phase I via modulated SN feedback.

The procedure is as follows: We start the simulation with star formation enabled, but set the lifetime of stars to a tenth of their normal value. We show the gaseous disk on initial conditions in Fig.\ \ref{fig:preproc:flat}, left column. Additionally, we enable mass return with the SN, i.e. the SN do not just return energy or momentum to the surrounding gas, but with each exploding SN, the mass of the star particle $M_\mathrm{starP}$ is reduced by $M_\mathrm{starP}/N_\mathrm{SN, tot}$, where $N_\mathrm{SN, tot}$ is the total number of SNe going off in the star particle, i.e. the number of stars with a mass $>8\,\mathrm{M}_\odot$ at the formation time of the star particle. The mass each SN returns is distributed evenly to the cells within the injection radius.

In this phase, the stars in the star particles explode as SNe relatively rapidly after their formation, and the mass return ensures that the star particles are gone when all stars are exploded, i.e.\ that no mass is locked up in the star particles. The energy and momentum injection by the SN induce turbulence in the disk.

Over the course of 1~Gyr we increase the lifetime of the stars in newly formed star particles back to their normal, tabulated lifetime. 
After reaching the normal lifetime, we run phase I for another Gyr, resulting in a total time in phase I of 2~Gyr. We show the result of this phase I in the Appendix, Fig.\ \ref{fig:preproc:flat}.

After this initial phase I, we follow phase II with regular stellar lifetimes and without mass return during SN events. In this paper, we only analyse phase II.

\section{Morphology}
\label{sec:results:morphology}

The Galaxy is a non-linear and highly complex system, with many processes at play. The gravitational potential of old stars and DM, which we here call the external potential, is only one of them. In this section, we study how the details of the Milky-Way external potential influence the morphology of the gas and the newly created stars. We are interested in the large scale structures, such as the bar and the spiral arm, as well as the influence of the vertical scale height of the disk.

\begin{figure*}
    \centering
    \includegraphics[width=\textwidth]{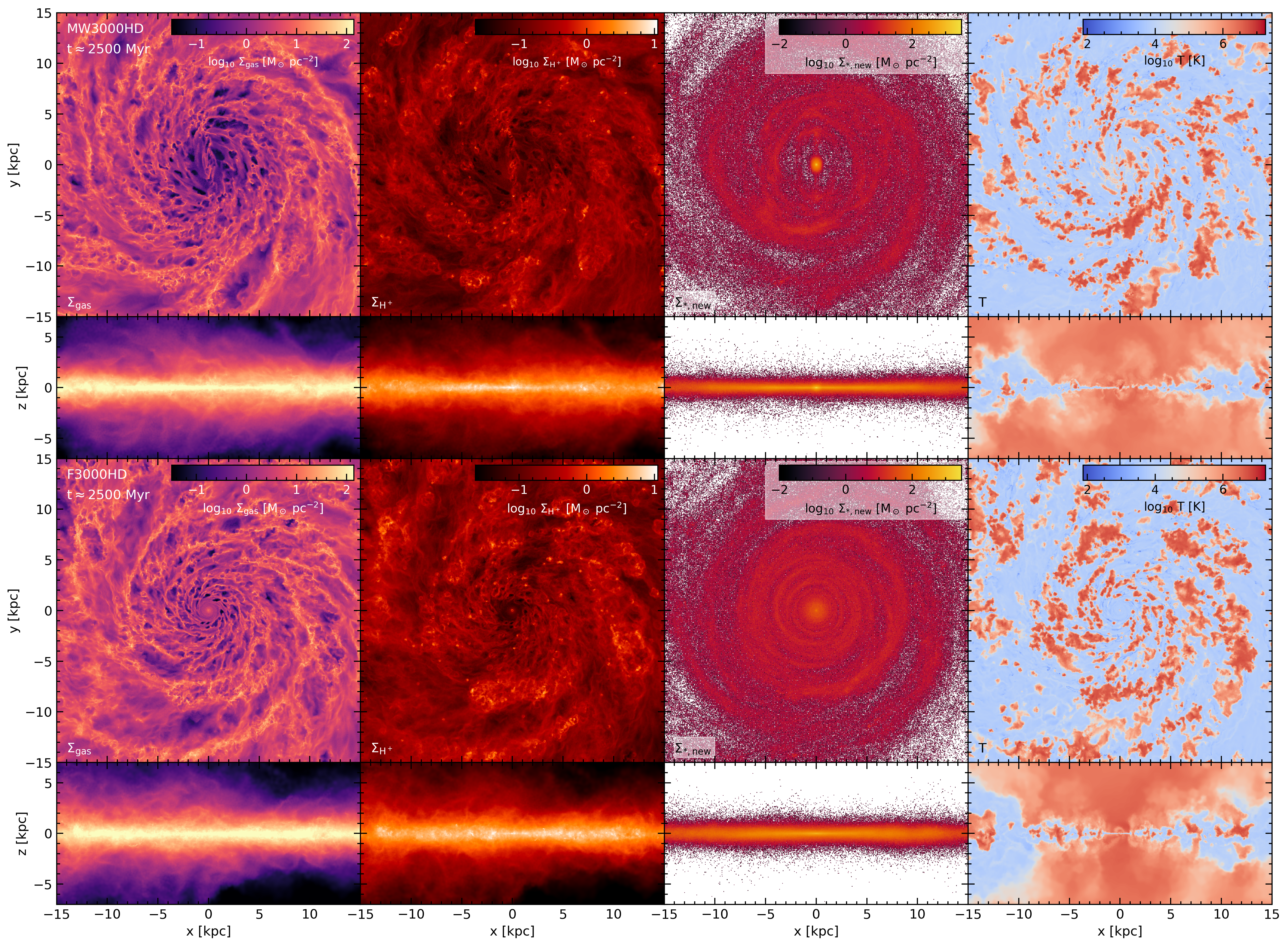}
    \caption{Surface density of total gas (first column), ionized hydrogen (second column), stars (third column) and the temperature in a slice through the midplane (fourth column) for simulations MW3000HD (first row) and F3000HD (second row) for the inner 15~kpc.}
    \label{fig:morphology_plot_center}
\end{figure*}

\subsection{Overall morphology}
\changeSummary{{Former Fig. 3 was removed and MHD simulations were removed from the remaining figure, the text is refined accordingly.}}

The external potential influence dramatically the overall aspect of our simulated galaxies.
The overall morphology of the galaxies differs early on, as shown by the column density at the end of phase~I in the appendix, Fig.~\ref{fig:preproc:flat}. 
Here, we focus on the morphology at our fiducial analysis time of $t=2500\,\mathrm{Myr}$, which is depicted in Fig.~\ref{fig:morphology_plot_center} (a depiction of the whole Galactic disk can be found in Fig. \ref{fig:morphology_plot}). From left to right we illustrate the surface densities of the total gas, the ionized gas, the stars formed during the simulation as well as the temperature in a cut through the midplane. In each panel, we show the face-on view of the disk as well as the edge-on view below for MW300HD (top) and F3000HD (bottom) (see Table \ref{tab:simulations}). The contribution of the bar potential results in a stronger inflow of gas into the central region ($R\lesssim 5\,\mathrm{kpc}$). This is visible in a bar-like structure of about 5~kpc length in the Galactic center in both $\Sigma_*$ as well as $\Sigma_\mathrm{gas}$. This results in a stronger concentration of mass in the center and a higher central star formation rate leading to a bulge-like structure visible in the edge-on $\Sigma_*$ projection. The regions outside the central zone are only marginally affected. The gas-free region around the Galactic center is a `depleted' area due to the bar. The bar has channeled all of the gas in this region towards the center. In our interpretation, it is part of the gap that forms around the Lindblad resonance (\citealt{SormaniSobacchiSanders2024}, see also \citealt{QuerejetaEtAl2021}).

We see the emergence of spiral arms in both the  gas and the stellar components for both potentials, which visually appear to be very similar. Morphologically, it is difficult to distinguish between spiral arms induced by a specific spiral component of the potential, and those that spontaneously arise in simulations due to the disk self-gravity and feedback from star formation.

The temperature slices show hot regions of recent SN activity distributed in spiral patterns following the stellar spiral arms in both simulations. The simulation MW3000HD shows those hot bubbles also in the center ($R_\mathrm{gal}\leq2.5$~kpc), whereas the central region of F3000HD is more quiescent. In the edge-on slice, hot outflows from the Galactic center are visible for both galaxies.

In summary, the inclusion of a barred external potential dramatically change the central region, with the formation of an otherwise absent bar shaped structure of gas and stars and an increased effect of the supernova feedback. In the rest of the disk, spiral structures forms through self-gravity even if it is not imposed by the gravitational potential. However we shall see in the following that these structures are transient, unlike the one created via an external potential. 

\subsection{Thermodynamic properties}
\label{sec:temp-dens-struct}
\changeSummary{{MHD simulations were removed, analysis of the phaseplots was adjusted to contain informations specific to the disk and center of the simulated galaxies, as differences between potentials emerge there. Fig. 7 and unnecessary references to different gas phases were removed.}}

\begin{figure*}
    \includegraphics[width=\textwidth]{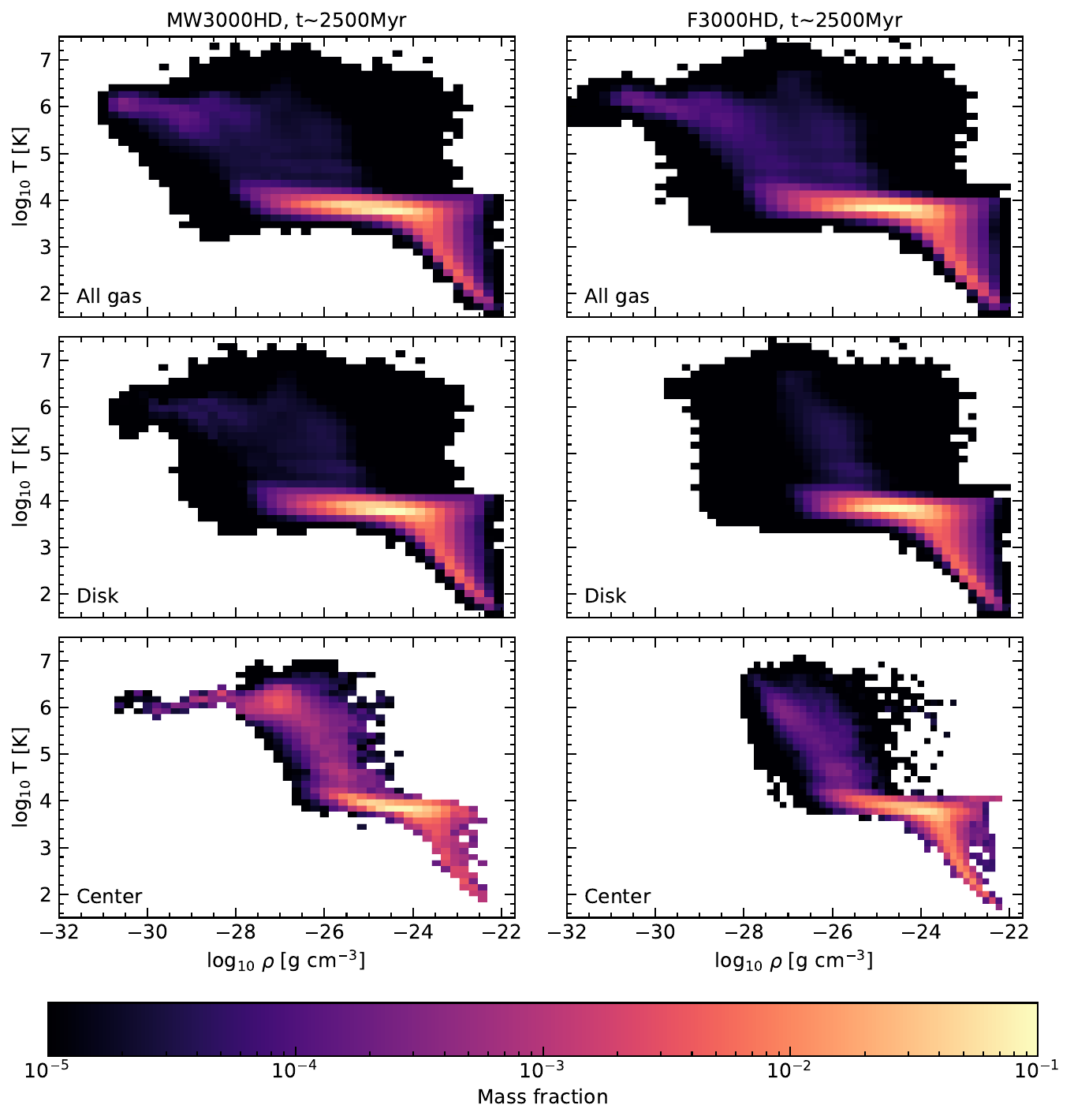}
    \caption{Temperature-density plot for the conducted simulations at the fiducial time of 2500\,Myr. All phase plots show a three-phase structure of the ISM, with a hot, low-density phase and a colder phase extending to higher densities, and the beginning of a cold phase at densities $\rho>10^{-24}$~g~cm$^{-3}$. Some mass is also stored in hot, higher-density gas, probably because of heating by stellar feedback.}
    \label{fig:phaseplot}
\end{figure*}

We have seen that the addition of a barred external potential changes the density structure of the gas in the center and the amount of supernovas exploding there. In this section we look at how the affect the phase of the gas in the different regions of the Galaxy.

In Fig.\ \ref{fig:phaseplot} we show the temperature-density distribution of the conducted simulations for the whole simulation box (top row), the Galactic disk ($|z|<1$~kpc and $R_\mathrm{gal}<25$~kpc, middle row) and the Galactic center ($|z|<1$~kpc and $R_\mathrm{gal}<2.5$~kpc, bottom row). Colour coded is the mass fraction of the respective part of the simulation.

The phase plots of the full simulation box show remarkably similar features between simulations: the main bulk of the gas resides in the area of the warm neutral and ionized medium at around $\mathrm{T}\sim 10^4$\,K, and, at densities larger than $10^{-24}$\,g\,cm$^{-3}$, extends into the cold neutral medium. At densities $\lesssim10^{-28}$~g~cm$^{-3}$, gas can only be found in the hot phase, at about $10^{5.5}-10^7$\,K. However, all simulations show hot gas also at densities of $10^{-28}-10^{-23}$~g~cm$^{-3}$. This is gas recently heated by SN feedback at its density during the energy injection, which has not yet expanded.
When zooming in onto the Galactic disk, the two simulations barely differ. Most of the mass in the hot gas is located in the circum-galactic medium, which explains the small amount of hot low-density gas. Conversely, the amount of gas in the warm and cold phase is hardly reduced compared to the full simulation box, meaning that this gas is mainly found within the Galactic disk.
When looking only at the Galactic center, for both simulations most gas is found in the warm phase. For F3000HD, more gas can be found in the cold phase than for MW3000HD, where instead some hot and low-density gas is present. This is a result of higher star formation and SN rates and denser clustering of SF and SN in the Galactic center, as we will see in Sections \ref{sec:rad_sf}, \ref{sec:cluster_sf} and \ref{sec:cluster_sn}.

\subsection{Global dynamical effects of the Galactic potential}
\changeSummary{{Former Fig. 1 and Fig. 5 were streamlined and merged. The corresponding sections were merged and rewritten to be more focused. Second panel of former Fig. 8 got removed, the text was shortened. The parts about the potentials and the velocity curve were merged}}

\changeSummary{{}}

\begin{figure}
    \includegraphics[width=\linewidth]{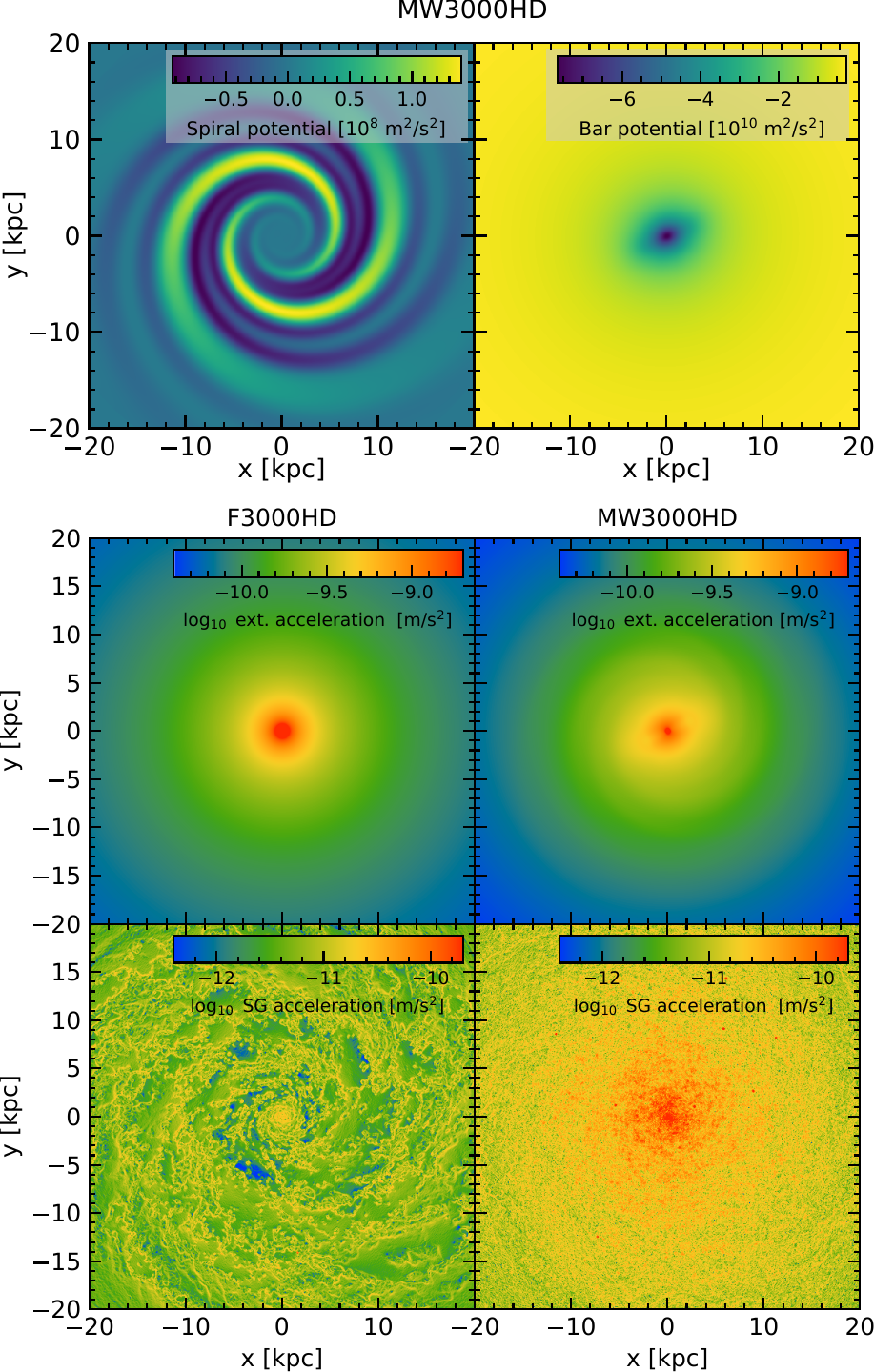}
    \caption{Gravitational potential and acceleration.
    Top: Non-axisymmetric component of the MW external potential. Bottom: Slices through the center of the galaxy of the acceleration parallel to the Galactic plane in simulations F3000HD (left column) and MW3000HD (right column) at $t\approx2500$~Myr.
    The first row corresponds to the contribution to the acceleration from the external potential, and the second is the contribution from the self-gravity (SG) of the gas and the newly created stars.}
    \label{fig:potentials}
\end{figure}

\begin{figure}
    \centering
    \includegraphics[width=0.49\textwidth]{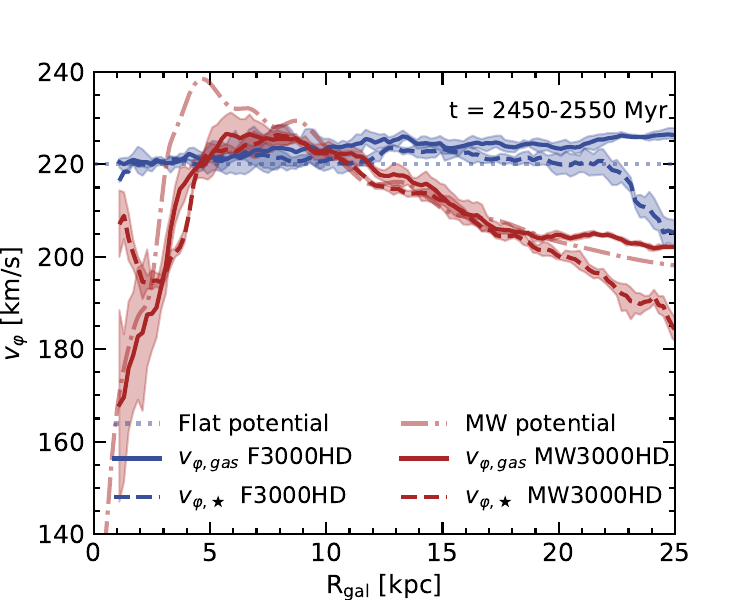}
    \caption{Azimuthal velocity mass-averaged over radial bins of 200\,pc averaged from 2450 to 2550~Myr for the gas (solid line) and the newly-formed stars (dashed line) with $\vert z \vert < 50$\,pc for the simulation with the MW external potential (red) and the flat external potential (blue). 16th to 84th percentiles are given as shaded regions. The equilibrium velocity curve that balances the radial force associated with the external potential is depicted by dash-dotted lines.}
    \label{fig:velocity_curve}
\end{figure}

In order to better understand how the external potential impacts the morphology of the galaxy, we give more details about the external potential we use. We show how they directly impact the dynamics by looking at the gravitational acceleration and the resulting velocity curves.

In our simulation, the gravitational potential is the sum of an external potential (accounting for the influence of the old stars and dark matter) and a contribution of the gravity from the gas and the newly-formed stars (through star particles). We remind that we compare the effect of a `flat' axisymmetric external potential with a flat velocity curve, and a multi-component potential tailored to mimic the Milky Way \citep{Glen} including the bar and four spiral arms (see Fig. \ref{fig:potentials}, top). We note that the strength of the spiral pattern is much weaker than the contribution of the bar.
In Fig.~\ref{fig:potentials} (bottom), we illustrate the corresponding accelerations for the two potentials at a time of 2500~Myr in a cut through the midplane. In the top row, we present the contribution due to the external potential; the bottom row shows the contribution from self-gravity. Both for the flat and the MW simulation the external potential largely dominates and is the main contribution to the full potential. The contribution of the bar is clearly visible in the visualization of the Milky Way potential. The spiral arm features, whose contribution to the potential is two orders of magnitude weaker, are vaguely visible (and therefore shown separately in the upper panel). In particular, in the simulation MW3000HD the contribution from the four spiral arms is outlined by a slightly higher acceleration than the background.
When we subtract the external potential from the total potential, we obtain the contribution from self-gravity of the newly formed stars and the gas (bottom row). The spiral structure is not present in the F3000HD simulation in the external potential but the self-gravity acceleration map clearly reflects the filamentary distribution of the gas and the stars. Interestingly, for MW3000HD, the acceleration arising from the self-gravity of the gas and newly formed stars is similar in shape to that of the external potential, with the self-gravitating acceleration being two order of magnitude lower. This is due to the stellar distribution closely following the potential wells generated by the bar and spiral arms, as discussed in section~\ref{sec:azimuthal}.

Figure~\ref{fig:velocity_curve} shows the rotation curve averaged over the time window from 2450-2550~Myr for both simulations in stars (dashed) and gas (solid lines). In dotted and dashed-dotted lines, we depict the equilibrium velocity needed to balance the radial force that is exerted by the external potential. For F3000HD, this velocity is 220~km~s$^{-1}$ throughout the whole disk, whereas for the MW potential it peaks at nearly 240~km~s$^{-1}$ at about 5~kpc and decreases to $\sim200$~km~s$^{-1}$ at 25~kpc. In both cases the simulations broadly follow the equilibrium curves. 
As designed, F3000HD has a flat velocity curve throughout the whole Galactic disk in both, stars and gas. The deviations from the equilibrium are minor (a few per cent), except for at large radii ($R_\mathrm{gal}\gtrsim20\,\mathrm{kpc}$), where the stars are below the equilibrium velocity and the gas is above it. 
In MW3000HD, the velocity curves of stars and gas are very similar to each other, but differ from the equilibrium velocity. In the inner galaxy,
up to $R_{\rm gal}$\,$=$\,12\,kpc,
where the equilibrium velocity is $>220$~km~s$^{-1}$, stars and gas have a velocity smaller than the equilibrium value, whereas further out, where the equilibrium velocity is $<220$~km~s$^{-1}$, gas rotates faster than the equilibrium velocity. This, in principle, would lead to a rupture of the galaxy, with a net inward flow in the inner region and a net outward flow in the outer region. However, the equilibrium velocity is derived from the \textit{external} potential. In the simulations we get additional gravitational feedback from the self-gravity of the simulated disk, smoothing out these discrepancies and leading to a stable Galactic disk.

\subsection{Radial structure}
\changeSummary{We split former section 3.2.2 into two sections, dealing with the radial and vertical structure separately. Former Fig. 10 was compacted, Fig. 11 and 12 removed. The text is shortened and streamlined to compare changes in stellar and gas distribution induced by a change in potential.}

Given the very different velocity curves (Fig.~\ref{fig:velocity_curve}) and the different overall morphology of our simulated galaxies, one may think that the radially averaged distribution of the gas depends on the proper inclusion of all the components of the Galactic potential. We see here that this is not the case, and that, except in the central region, the flat external potential reproduces fairly well the radial structure of the galaxy modeled with a more complete potential.

In Fig.\ \ref{fig:dens_dist_general} we present the radial (top row) volume-weighted distribution of gas (left) and newly formed stars (right), averaged over 100~Myr around our fiducial analysis time of 2500~Myr. First, we point out that the density distribution only changes minimally during phase II as we reach a steady state.
Second, the radial distribution of the gas is very similar in all of the simulations: a relatively flat distribution, decreasing from $\sim10^{-24}$~g~cm$^{-3}$ close to the center to $\sim10^{-25}$~g~cm$^{-3}$ at the edge of the disk. Gas at these densities is predominantly warm ($T \sim 10^{4}$~K; see Figure~\ref{fig:phaseplot}), so in all simulations, the volume of the disk is filled primarily with warm neutral and warm ionized gas. The radial distribution of stars is significantly steeper than the gas distribution but is independent of the choice of potential. Within the disk, the density of stars steadily declines towards the Galactic outskirts, with a knee short of 25~kpc and a following, steeper decline (not shown here). We note that there is a noticeable difference between the flat and MW model at a galactocentric distance of $1-3\,\mathrm{kpc}$, both in the gas as well in the stars. This is due to the stronger star formation in the MW model, which manifests in a dip in gas and stellar density at $\sim2\,\mathrm{kpc}$, because the center depletes star forming material from that zone, and a central peak ($<1\,\mathrm{kpc}$) in the distribution of stars.

We bring to the reader's attention that the depicted curves do not represent the full stellar population present in the simulation boxes, but the one formed after the simulation started (in phase II). The external potential applied to the simulated galaxies provide an additional mass distribution, mainly from (low-mass) stars and DM. For the MW potential (grey dash-dotted curve), it includes only the stellar and gaseous component, for comparability (if DM would be included, it would dominate over the potential stellar and gaseous component at galactocentric radii $R_{\rm gal}$\,$>$\,$8.7$~kpc). For the flat potential (grey dotted curve), it is not possible to distinguish between gaseous, stellar and DM components, since it is an analytical potential which describes an overall density distribution. We therefore plot the distribution of all underlying mass, including possible DM.

\subsection{Vertical structure}
\changeSummary{{Same as for section 3.4.}}

\begin{figure*}
\centering
    \includegraphics[height=7.5cm]{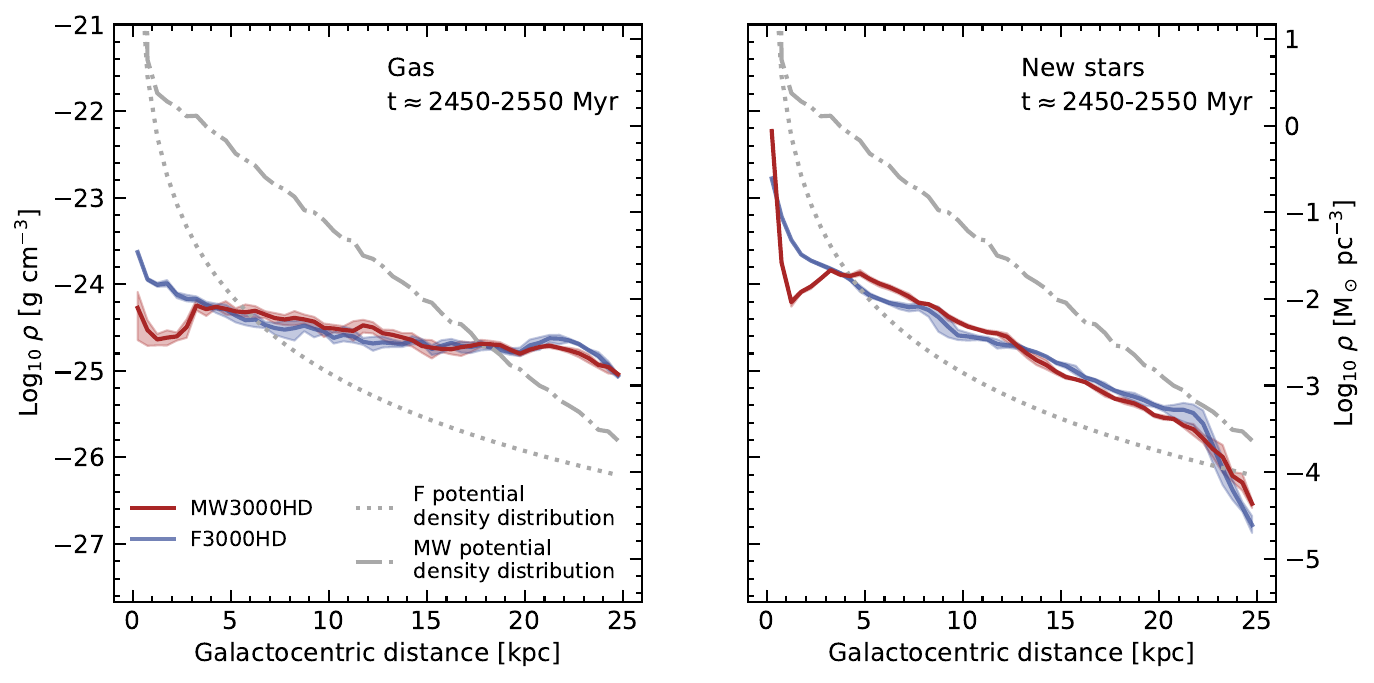}\\
    \includegraphics[height=7.5cm]{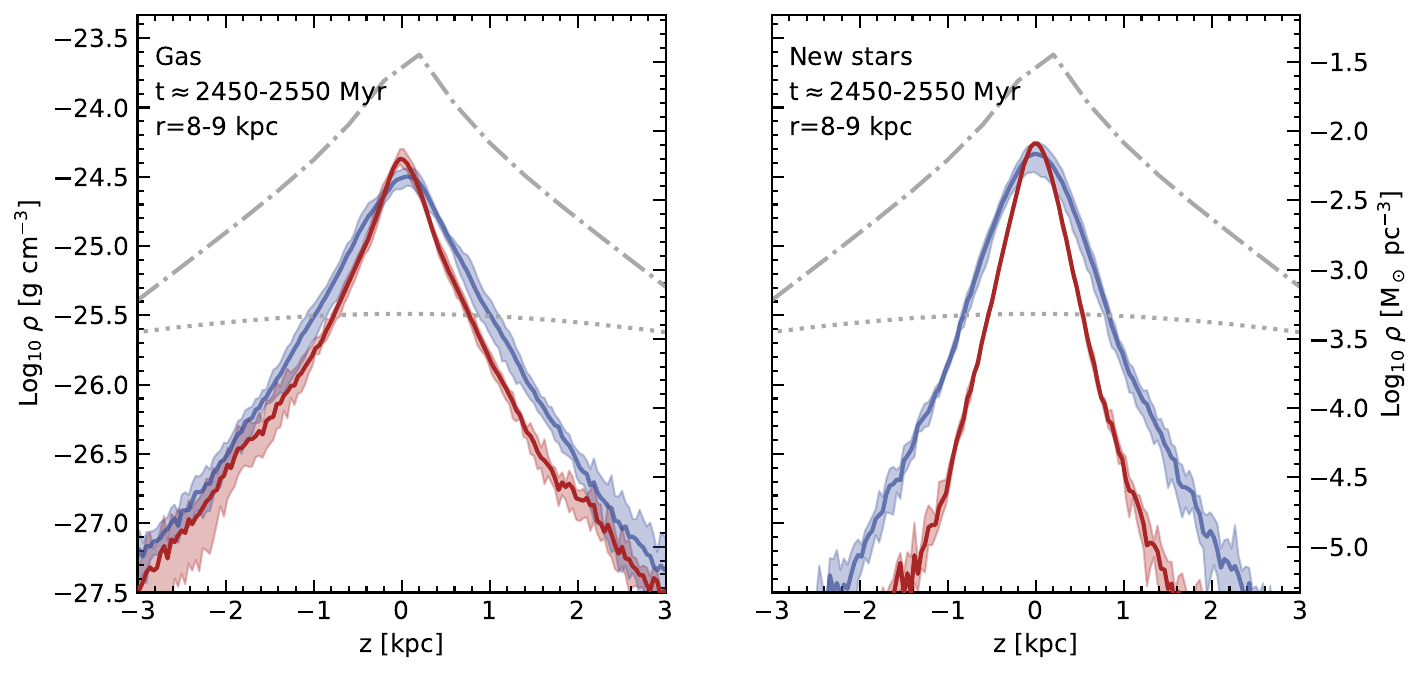}
    \caption{Volume-weighted density distribution of gas (left) and stars (right) in the radial (upper row) and vertical (lower row, averaged from 8 to 9~kpc) directions for F3000HD (blue) and MW3000HD (red) 
    averaged from 2450 to 2550 Myr. Shaded regions indicate 16th to 84th percentile. The density distributions generating our adopted external potentials are indicated in grey, dotted for the flat potential and dash-dotted for the Milky Way model. In the radial direction, we take into account mass up to $z=\pm50$~pc.}
    \label{fig:dens_dist_general}
\end{figure*}

 The external potential is also expected to change the vertical equilibrium through the disk.
 We look at it in the bottom panel of Fig.\ \ref{fig:dens_dist_general}. Again, we find that the vertical distribution of gas density is nearly identical between the two simulations. The stars however are not sensible to pressure gradients. The vertical stellar distribution for MW3000HD is narrower than for F3000HD, i.e.\ stars are more concentrated towards the Galactic midplane. 

\begin{figure*}
    \centering
    \includegraphics[width=\textwidth]{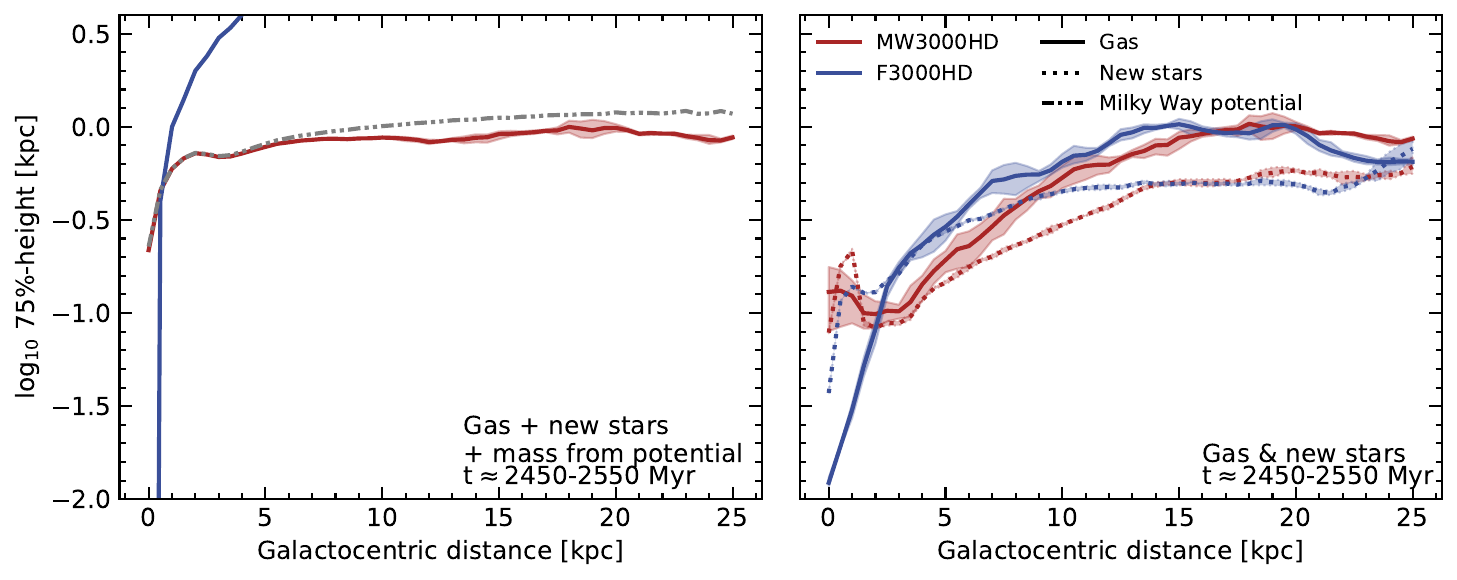}
    \caption{Height of the plane containing 75\% of the mass at a given radius, for all simulated mass (left), simulated gas and newly formed stars (right) in MW3000HD (red) and  F3000HD (blue). In grey we present height for stars in the MW potential, i.e.\ the underlying, not explicitly simulated mass distribution. Shaded regions indicate 16th to 84th percentile. In all simulations, stars are more concentrated to the disk plane than gas.}
    \label{fig:scaleheights}
\end{figure*}

We complete this analysis by considering the scale height of the gas and stars, which we define as the  height containing 75\% of the mass at a given radius. We call this height `75\%-height'. Again, we average over a timespan of 100~Myr, from 2450 to 2550~Myr. On the left in Fig.\ \ref{fig:scaleheights} we show the 75\%-height for all mass in the simulation, i.e.\ gas, newly-formed stars and mass in the external potential. There, in grey we also present the height of the stars in the MW potential, i.e.\ the underlying, not explicitly simulated mass distribution. On the right we show the 75\%-height for the gas and newly formed stars in the simulation.

The mass distribution of the external potential completely dominates the 75\%-height for F3000HD (see left panel). The underlying mass distribution of the flat potential is not shown, but identical with the blue solid curve. This is, because the flat potential, as mentioned before, is an analytical potential, and we cannot distinguish between the DM and other components. Therefore, we use the full mass distribution forming the external potential in this figure. As DM is a fuzzy component, it drives the 75\%-height to higher values. For the MW potential we remove the DM component for this comparison. MW3000HD is dominated in its 75\%-height by the underlying potential mass as well (compare grey dash-dot-dotted curve vs solid red curve) up to about 5~kpc. At higher galactocentric radii, however, the simulated mass components lower the 75\%-height, i.e. they are more closely concentrated to the Galactic disk plane than the mass causing the potential.

In stars and gas, a general trend is present in both simulations of increasing 75\%-height with increasing $R_{\rm gal}$ up to a radius of about 10-15~kpc, with a saturation at higher galactocentric radii. For gas, the saturation is about 1~kpc; for stars it is slightly lower, at about 0.5~kpc for both potentials. Stars at $R_\mathrm{gal}\gtrsim10$~kpc have a generally lower 75\%-height than the gaseous component, meaning they are confined closer to the disk than gas.
Within the innermost $\sim2$~kpc, the 75\%-height of stars shows a bump, caused by local turbulence from close interaction in this region. 
This, however, is more pronounced in the MW potential, reaching about $\sim 0.5$~dex in height, and is most probably caused by turbulence induced by the bar and the lower potential strength compared to the flat potential (see Fig.\ \ref{fig:potentials}).

From Eq. \ref{equ:density_profile} it is easy to show that the 75\%-height $x_{75\%}$ is about twice the scale height $z_d$ by $\int^{x_{75\%}}_{-x_{75\%}}\rho(\Sigma, z)dz=0.75\int^\infty_{-\infty}\rho(\Sigma, z)dz$ which results in $x_{75\%}\approx1.94z_d$. The scale height therefore is about 500~pc in gas and 250~pc in stars for MW3000HD. The stellar scale height is at the same order as the observed scale height of the thin stellar disk (which the young stars formed during our simulation represent) of $279.76\pm12.49$~pc \citep[from Gaia D3,][]{scaleheight_stars}. However, our simulated galaxy is dynamically young and lacks interaction with satellites; the stellar disk would probably thicken if run for a longer time. \citet{Ferrire2001} quotes (exponential) scale heights up to 1000~pc (WIM) for Galactic gas (which corresponds to about 900~pc in $\mathrm{sech}^2$). Our found scale height falls well within this limit.

\subsection{Azimuthal structure}
\label{sec:azimuthal}
\changeSummary{{Former Fig. 13 got reworked, the main focus is now on the last row of this figure, which we discuss in Section \ref{sec:az_sf}.}}

\begin{figure*}
    \centering
    \includegraphics[width=0.95\textwidth]{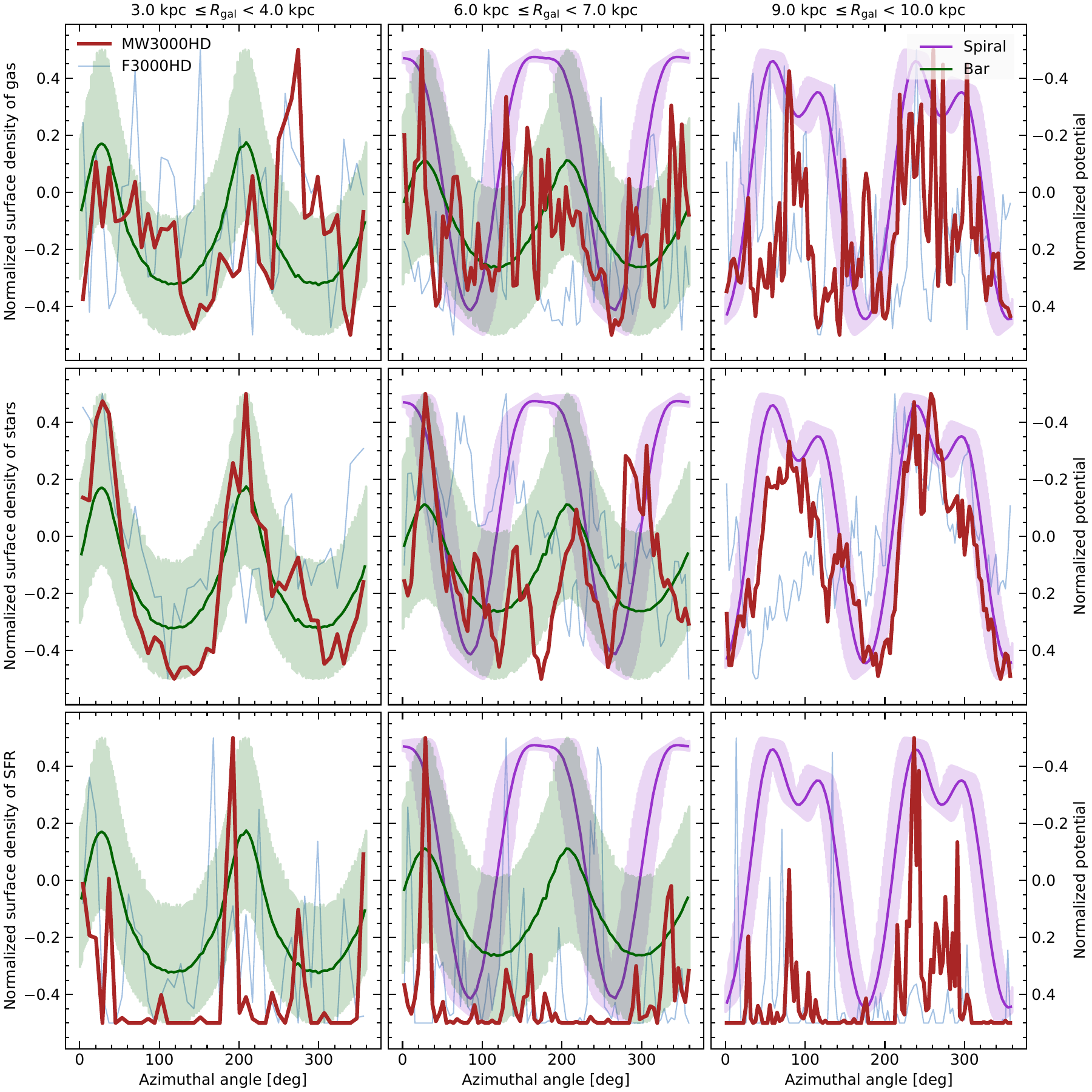}
   
    \caption{Azimuthal profile of the normalized fluctuations of the gas (top), stars (middle) and SFR (bottom) surface densities at t = 2500 Myr. They are compared with the normalized fluctuations of the contributions of the external non-axisymmetric components of the potential, with the bar (green) and  the spirals (pink), associated to the right vertical axis of each panel, reversed so that lower values are on top. The data is averaged over a radius range of 1 kpc and the shaded area represent the radial variations. }
    \label{fig:azprof}
\end{figure*}

The MW external potential is characterized by the presence of two non-axisymmetrical components, the bar and the spiral arm. To have a better idea on how they shape the galaxy, we look at the repartition of the gas and stars in the azymuthal direction at different radii.

Figure~\ref{fig:azprof} shows the azimuthal profile of gas (top panel) and stars (middle panel) at 3\ -\ 4~kpc, 6\ -\ 7~kpc and 9\ -\ 10~kpc. We compare the profile for the simulations F3000HD and MW3000HD. 
We also plot the contribution of the dominant external non-axisymmetric components of the MW3000HD potential for reference in green and pink.
In this plot we focus on the azimuthal fluctuations of each quantities. Each quantity $y$ is averaged in azimuthal bins and normalised as follows:
\begin{equation*}
    y_\mathrm{norm} = \dfrac{y- (y_\mathrm{max} + y_\mathrm{min})/2 }{y_\mathrm{max}  - y_\mathrm{min} }
\end{equation*}
As a result, the plotted value $y_\mathrm{norm}$ is  dimensionless and varies between $-0.5$ and $0.5$.

At all radii, the simulation F3000HD shows no clear large-scale structure, with many peaks in the surface density profiles.
On the contrary, the simulation with the MW  potential shows fewer peaks which are well correlated with the bar and the spiral arms. 
The best correlation found are for the stars, where they follow the potential minima of the bar (middle left panel) or the spiral arms (middle right panel). What happens around 6 kpc (middle center panel) is less clear, as the influence of the bar and the spiral arm are combined.

Overall, our analysis shows that choosing an accurate prescription of the Milky-Way potential does not have a big influence on the radial distribution of the gas and young stars, except in the central regions where the bar is prominent. 
However, the analysis reveal that the bar and the spiral arms have a significant influence on the organization of the gas and stars in the inner and outer galaxy when looking in the azimuthal direction.

\section{Star formation and stellar feedback} \label{sec:results:sf}
\changeSummary{{This section got expanded as we focus on differences in star formation induced by a change in Galactic potential.}}

In the previous section, we analysed the effect of an accurate prescription of the MW potential on the morphology of the galaxies, focusing on the gas and star structures. Quite naturally, we expect that a change of the morphology of the gas reflects into a change in where the stars form and the overall star formation properties. In this section, we inspect what is the effect of changing the external potential on the star formation activity, starting from a global point of view and then again refining in the radial and azymuthal directions.

\subsection{Global star formation rate is agnostic of potential}
\label{sec:global_sf}
\changeSummary{{MHD simulations were removed from former Fig. 14, the text got remodeled to better explain the extraction of SFR from the simulations, and expanded for a more detailed analysis of the depletion time.}}

\begin{figure*}
    \centering
    \includegraphics[width=0.8\textwidth]{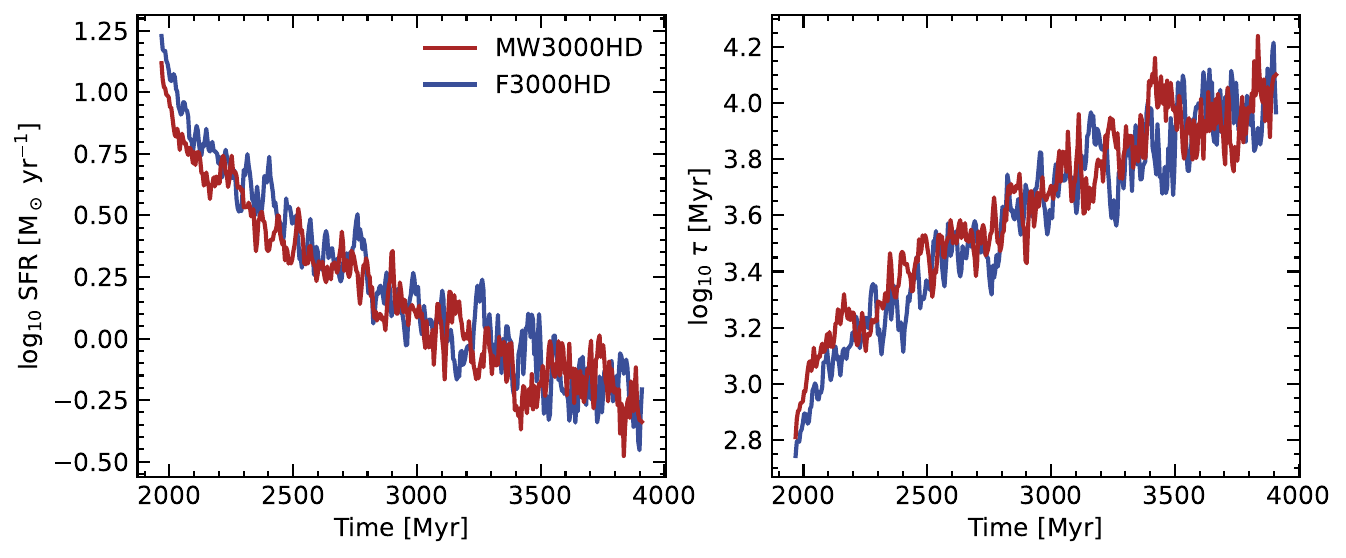}
    \caption{SFR history (left) and depletion time $\tau$ (right) for F3000HD (blue) and MW3000HD (red). We limit the measurement to $\pm1$~kpc around the Galactic plane. Changing the potential does not result in a change of the overall SFR at the fiducial analysis time.}
    \label{fig:SFR}
\end{figure*}

We start by looking at the total SFR in our simulations over time in Fig.\ \ref{fig:SFR} (left) as well as the gas depletion time (right). We define the SFR as
\begin{equation}
    \mathrm{SFR}_n=\frac{M_*(t_{n+1})-M_*(t_n)}{t_{n+1}-t_n},
\end{equation}
with $t_n$ and $t_{n+1}$ being the times of two consecutive snapshots and $t_{n+1}-t_n \approx 5\,\mathrm{Myr}$ in our case. The gas depletion time, consequently, is defined by
\begin{equation}
    \tau=\frac{\bar{M}_\mathrm{gas}}{\mathrm{SFR}},
\end{equation}
where $\bar{M}_\mathrm{gas}$ is the arithmetic mean of the gas mass of two consecutive snapshots.

An initial peak is present in both simulations, which results from an initial collapse of the disk when turning off the mass return from supernovae at the end of phase I and the corresponding additional pressure.

Both simulations show very similar global SFRs, declining constantly over the simulation time, which is consistent with the increasing depletion of the gaseous disk by star formation in an isolated galaxy. At our fiducial analysis time, we find SFRs of 2.9~\uSFR\ (F3000HD) and 2.6~\uSFR\ (MW3000HD). These values are at the upper end of estimated SFRs from observations, which mostly suggest a global SFR for the Milky Way of about $1-3$~\uSFR\ \citep[see e.g.][and references therein]{Chomiuk2011, Licquia2015, Bland-Hawthorn2016, Elia2022}. At a simulation time of 3000~Myr, the SFR has decreased to about $1.5$~\uSFR\ (F3000HD), $1.3$~\uSFR\ (MW3000HD), which fits observational constraints rather well. From about 3500~Myr simulation time onward, SFR is below 1~\uSFR and therefore too low for a Milky-Way analogue. 

The depletion time $\tau$ increases during the simulation from less than $10^3$ to more than $10^4$~Myr. The molecular gas depletion time of low redshift main sequence galaxies is found to be between $900$ and $2000$~Myr \citep{WangTsan2022}. Our simulations exceed this range after about 2500~Myr. At later simulation times, it therefore seems that the star formation efficiency is too low compared to the real Milky Way, possibly because of lack of disturbance of the galaxy by satellites that trigger additional star formation, and because of the fact that we do not model circum-galactic gas that could replenish the Galaxy.

\subsection{Radial distribution of star formation} \label{sec:rad_sf}
\changeSummary{{MHD simulations removed, and we added a more detailed view of star formation in the Galactic center, as this is where the main differences between potentials emerge.}}

\begin{figure*}
        \centering
        \includegraphics[width=0.8\textwidth, trim={0, 1.3cm, 0, 0}, clip]{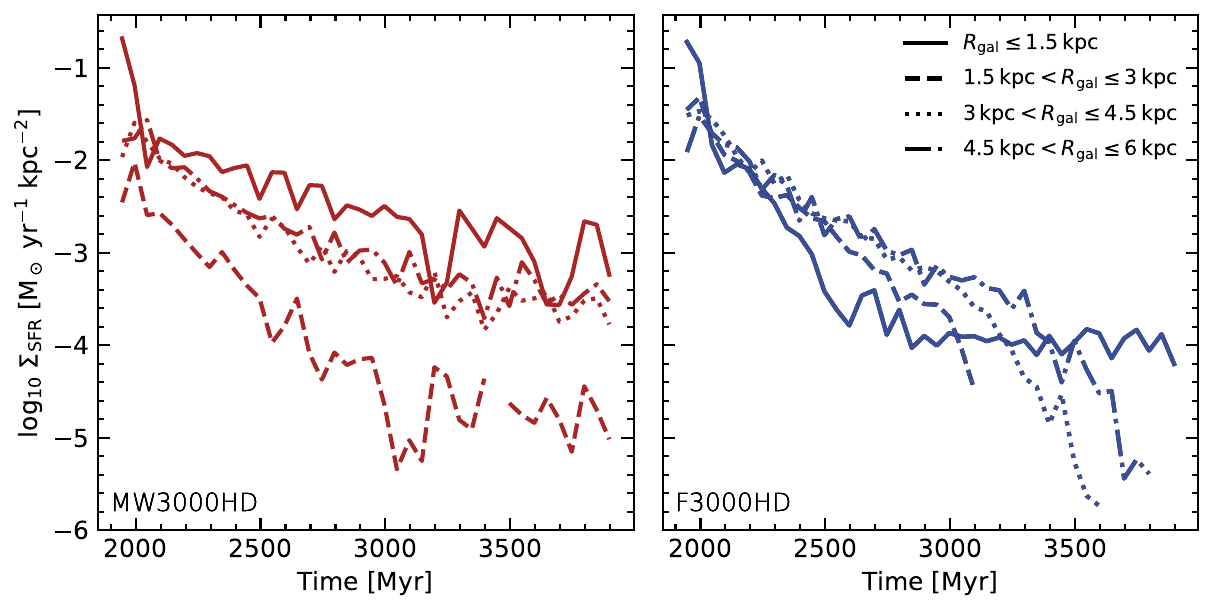}
        \includegraphics[width=0.8\textwidth]{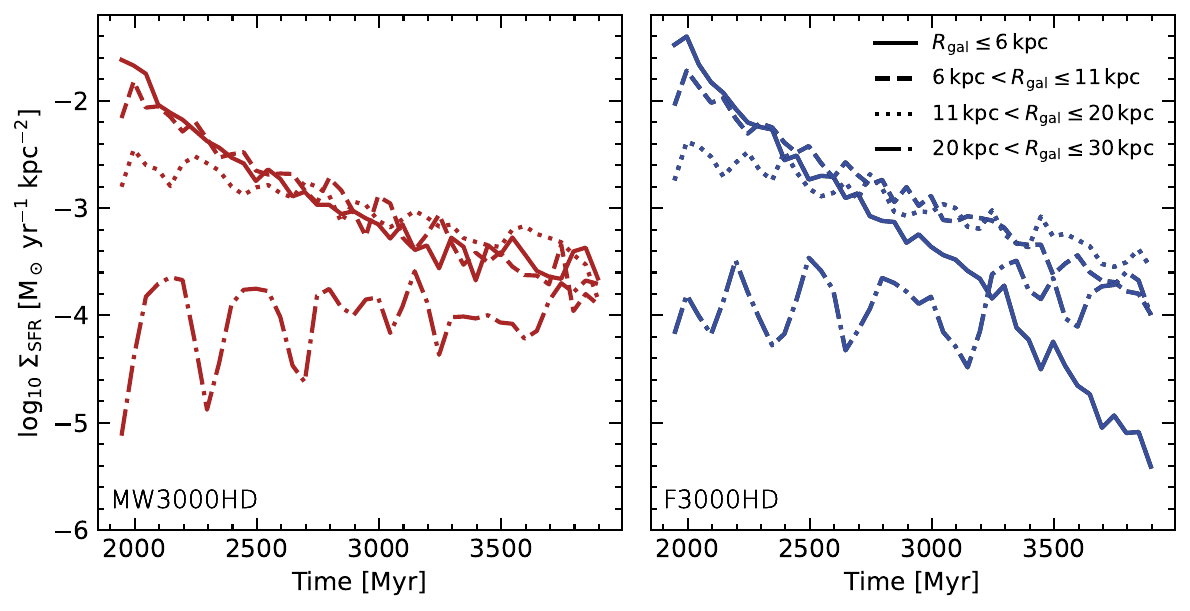}
        \caption{SFR surface density in MW3000HD and F3000HD in radial bins for the central ($R_\mathrm{gal}<6$~kpc) region (top row) and whole disk (bottom row).}
        \label{fig:SFR_radbins}
\end{figure*}

Since the global SFR is nearly identical for both potentials, in Fig.\ \ref{fig:SFR_radbins} we present the star formation rate surface density $\Sigma_\mathrm{SFR}$ in radial bins for the Galactic center (top row) and the whole disk (bottom row). Again, a constant decline in $\Sigma_\mathrm{SFR}$ is present at all radii, except the outermost bin of 20-30~kpc. The Galactic outskirts remain at a constant, but low $\Sigma_\mathrm{SFR}$. Comparing values for the whole disk in F3000HD and MW3000HD, the $\Sigma_\mathrm{SFR}$ at different radial bins largely agrees apart from the innermost 6~kpc. For F3000HD, star formation in this region declines steeply with time, with the SFR of the center falling below the average SFR of the rest of the disk about halfway through phase II. Conversely, in run MW3000HD, $\Sigma_\mathrm{SFR}$ does not decline below the levels of regions further out. This most probably is because of the constant channeling of gas into the Galactic center by the bar, which prevents quenching. Soon after the beginning of phase II, $\Sigma_\mathrm{SFR}$ is rather constant from 6 up to 20~kpc, i.e.\ the corresponding curves overlap to a large degree.

Zooming in onto the Galactic center (top row), differences in the distribution of star formation become more apparent. Whereas for F3000HD, star formation is mostly evenly distributed throughout the innermost 6~kpc ($\Sigma_\mathrm{SFR}$ agrees for different radial bins) and is lowest for the innermost 1.5~kpc, a clear trend in $\Sigma_\mathrm{SFR}$ is visible for MW3000HD. We find the highest $\Sigma_\mathrm{SFR}$ within the innermost 1.5~kpc. Between $1.5-3$ and $3-4.5$~kpc, $\Sigma_\mathrm{SFR}$ is roughly the same, with a value $\sim 0.5$~dex below the value in the innermost bin. Between $4.5$ and $6$~kpc, far fewer stars are formed than at smaller radii. After about 2500~Myr, $\Sigma_\mathrm{SFR}$ in this region is comparable to or lower than the SFR density at $R >20$~kpc, i.e.\ this zone is quenched. However, from $0$ to $4.5$~kpc star formation does not quench during the duration of the simulation. This is most probably prevented by constant gas inflow into the center due to gas dynamics in the bar. Fig. \ref{fig:SF_spirals}a shows this in more detail, as we will explain in Section \ref{sec:az_sf}

F3000HD on the other hand shows a severe drop-off and even complete stop of star formation after about 3000~Myr. Whereas $\Sigma_\mathrm{SFR}$ is lowest within $1.5$~kpc galactocentric radius for most of the simulation time, star formation does not fully stop there at any time, but continues on a low and constant level from 2500~Myr onwards. 

We therefore conclude that gas streaming due to bar dynamics has a large influence on star formation in the center of the galaxy.

\subsection{Azimuthal distribution of star formation}\label{sec:az_sf}
\changeSummary{{We added more detailed view on star formation differences between potential in the azimuthal direction. Most of this section is not contained in the former version of the paper.}}

While both simulation have spiral arms and show a similar star formation rate radial profile outside the central region, we have seen in section \ref{sec:azimuthal} that the MW external potential has a strong influence on the azimuthal distribution of the gas and stars. It is thus interesting to look at the azimuthal distribution of the SFR, which we do in the bottom row of Fig.\ \ref{fig:azprof}. Again we look at radial bins from 3-4~kpc (left), 6-7~kpc (middle) and 9-10~kpc (right), for F3000HD (blue) and MW3000HD (red), together with the MW potential for the bar (green) and the spiral arms (pink). The value of the SFR is also normalized as described in Section \ref{sec:azimuthal}.
As we can expect, the SFR in F3000HD shows no periodicity in the azimuthal direction at any given radius.
On the other hand, the SFR in MW3000HD correlates well with low points in the dominating potential pattern, be it the bar (see left panel) or the spiral arms (see right panel). Within those potential wells, SFR is increased. In regions with no strong dominance of one potential pattern over the other (central panel), SFR presents no clear correlation with one or the other of the given potential patterns. 

\begin{figure*}
    \centering
     \begin{subfigure}{0.31\textwidth}
        \centering
        \includegraphics[height=5.5cm, trim={0, 0, 2.35cm, 0}, clip]{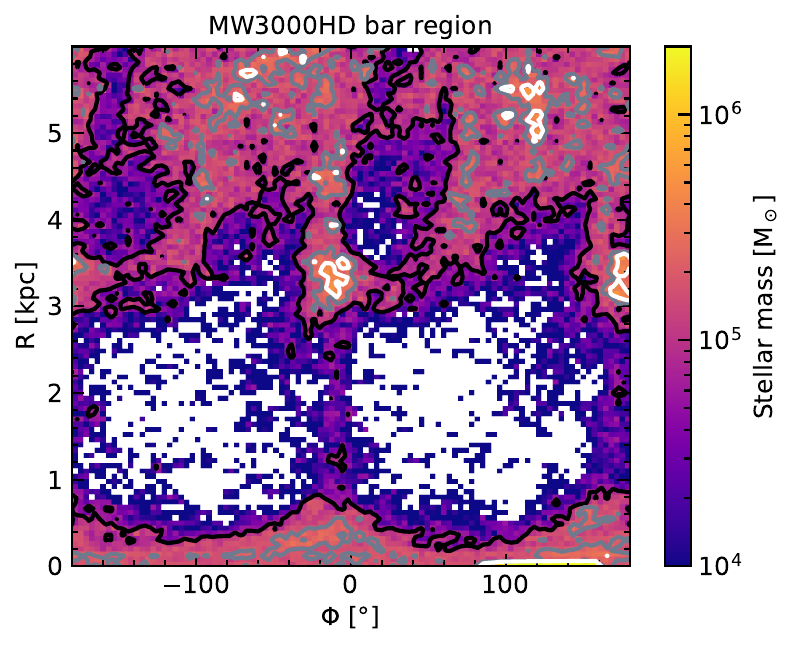}
    \end{subfigure}
    \begin{subfigure}{0.37\textwidth}
        \centering
        \includegraphics[height=5.5cm]{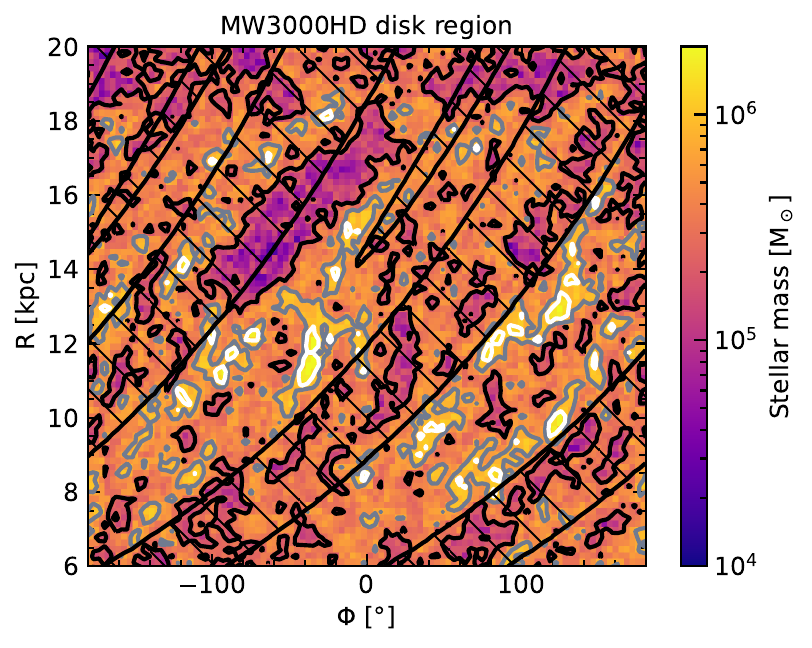}
    \end{subfigure}
    \begin{subfigure}{0.31\textwidth}
        \centering
        \includegraphics[height=5.5cm]{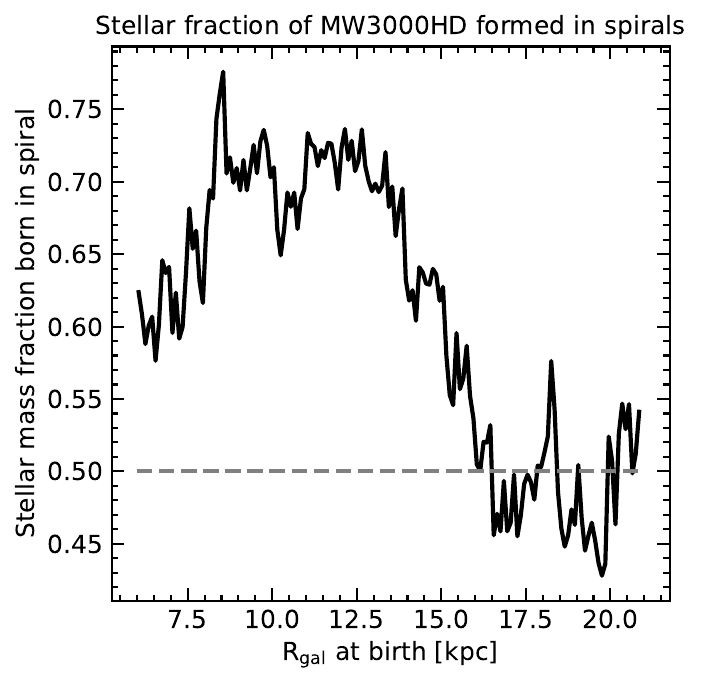}
    \end{subfigure}\\
    \begin{subfigure}{0.31\textwidth}
        \centering
        \includegraphics[height=5.5cm, trim={0, 0, 2.35cm, 0}, clip]{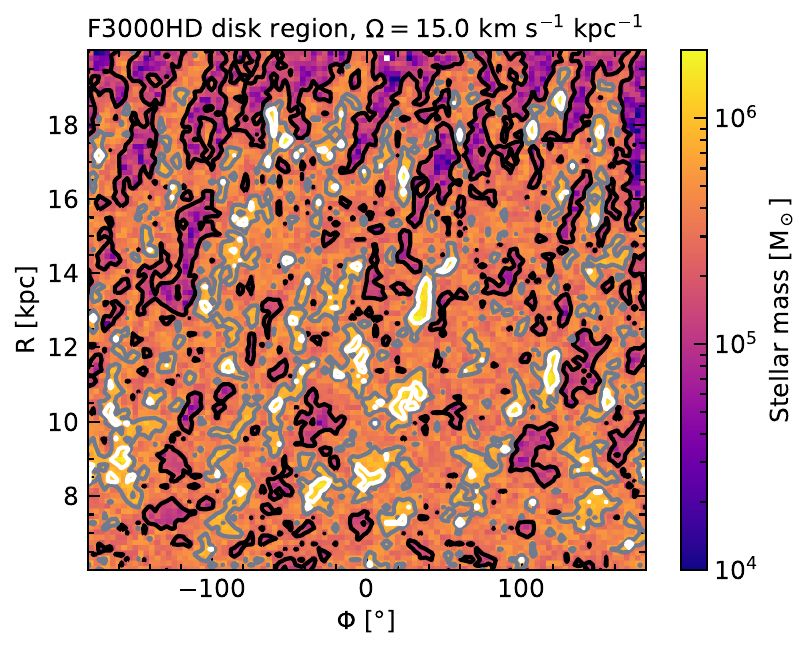}
    \end{subfigure}
    \begin{subfigure}{0.31\textwidth}
        \centering
        \includegraphics[height=5.5cm, trim={0, 0, 2.35cm, 0}, clip]{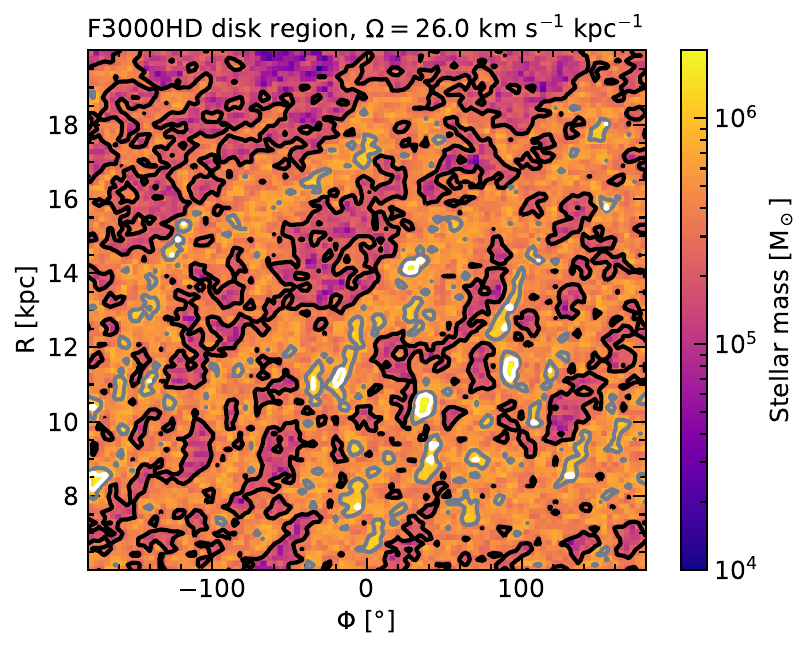}
    \end{subfigure}
    \begin{subfigure}{0.37\textwidth}
        \centering
        \includegraphics[height=5.5cm]{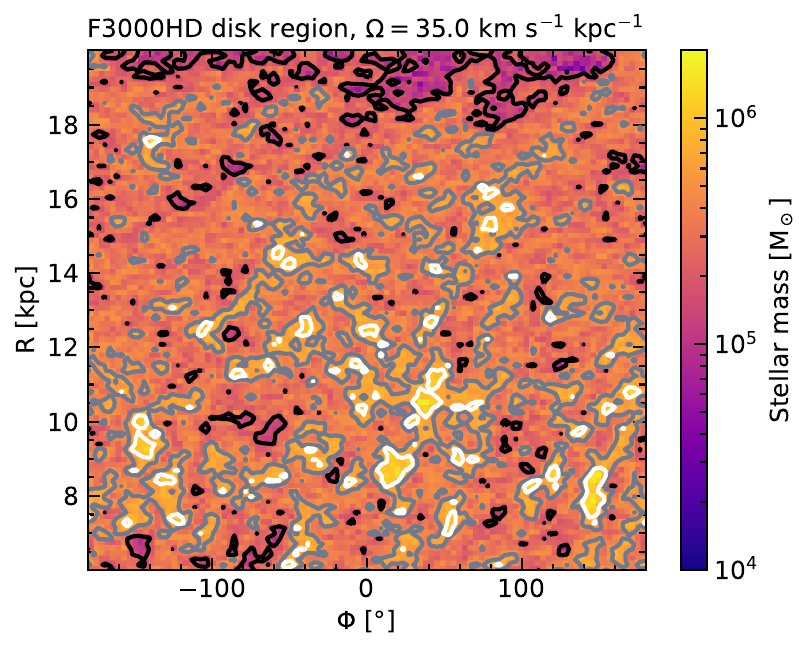}
    \end{subfigure}
    \caption{Stellar mass fraction formed in spirals in MW3000HD (top-right panel) and formed stellar mass in $R_\mathrm{gal}-\Phi$-projection for MW3000HD disk region ($6\,\mathrm{kpc}\leq R_\mathrm{gal}<20\,$~kpc, top-middle panel) and bar region ($R_\mathrm{gal}<6$~kpc, top-left panel), corrected for the corresponding pattern rotation, as well as for the disk region of F3000HD, corrected for a pattern speed of $\Omega=15.0$, $26.0$ and $35.0$~km~s$^{-1}$~kpc$^{-1}$ (bottom row). Contour lines indicate 10~\%, 30~\% and 50~\% of maximum stellar mass formed (excluding the region of extremely high formed stellar mass at $<0.1$~kpc for MW3000HD bar region, upper left). For MW3000HD disk region, upper middle, interarm regions are shaded out.}
    \label{fig:SF_spirals}
\end{figure*}

To expand this analysis and see if it is robust at different timestep, we need to look at where star formation occurs over time. To that purpose, we use rotation corrected $R$-$\Phi$-projection  indicating the total stellar mass formed in different regions of the galaxy. (Fig. \ref{fig:SF_spirals}). The figure spans the whole time of our simulation (phase II) and is corrected for rotation with the pattern speed of the Galactic bar (top-left panel) and spiral arms (top-middle panel). In the top-left panel  we do not see a clear pattern in where stars are forming for $R\gtrsim4.5$~kpc, as bar and spiral potentials with different pattern speeds overlap in this region (as already seen in Fig. \ref{fig:azprof}). However, between 1 and 3~kpc star formation is located within streams of gas to the Galactic center and circular active star formation zones at their tips (see contour lines). In the top-middle panels we show the disk region of MW3000HD, with non-spiral arm regions crossed out. For this purpose, we define spiral arms as regions where the non-axisymmetric spiral arm potential perturbation is lower than $0$~m$^2$~s$^{-2}$. As in Fig. \ref{fig:azprof}, we find that star formation occurs preferentially within the spiral arms. At radii $>14$~kpc, however, these star formation regions get out of sync with the inter-arm regions, because the rotational speed of the gas drops below the pattern speed (compare Fig. \ref{fig:velocity_curve}). 
This is also an evidence that the MW external potential yield long-lived spiral structure where star formation occurs preferentially.
To quantify this, in Fig. \ref{fig:SF_spirals} (upper right) we present the mass fraction of stars formed within spiral arms vs galactocentric distance. For $6\,\mathrm{kpc}\leq R_\mathrm{gal}\leq16$~kpc, more than half of all stellar mass is built within spiral arms, with a peak of 70-75\% between 9 and 14~kpc. Outside of 14~kpc, the fraction of stellar mass formed within spiral arms drops off steeply and reaches an equal distribution between arm and interarm regions at about 16~kpc. This, as mentioned before, is most likely caused by the discrepancy between the rotation speed of the gas, which slowly declines with radius (see Fig. \ref{fig:velocity_curve}), and the pattern speed of the spiral arm potential, which behaves like a solid body. In total, of all stars formed beyond 6~kpc, about 62~\% are formed in spiral arms.

Contrary to the Milky Way Model, for F3000HD (lower row of Fig. \ref{fig:SF_spirals}) we do not find similar noticeable features. Spiral structures do emerge in this simulation as well (see Fig.~\ref{fig:morphology_plot}) but they do not persist over time. 
Indeed, we produced a series of rotation-corrected $R$-$\Phi$-projection with angular frequency ranging from $15$ to $35.0$~km~s$^{-1}$~kpc$^{-1}$ in steps of $0.5$~km~s$^{-1}$~kpc$^{-1}$, of which three examples are shown ($\Omega=15,~26$ and $35$~km~s$^{-1}$~kpc$^{-1}$). We don't see any coherent spiral pattern in any of the tested angular frequency $\Omega$.

We have seen in section \ref{sec:global_sf} that the global star formation rate is similar in both sets of runs. That suggests that although differences in the potential clearly influence where stars form, they make little difference to the galaxy-averaged star formation efficiency. This is consistent with prior theoretical and observational work. For example, \cite{kimLocalSimulationsSpiral2020} simulated a local patch of the ISM with an imposed spiral potential and found that while star formation occurs preferentially in the spirals, where the gas is denser, the star formation efficiency is not enhanced by the spiral perturbation. On the observational side, comparisons of the star formation efficiency in spiral arms with the efficiency in inter-arm regions in our Milky Way \citep{Ragan2018} and other nearby spiral galaxies \citep{Sun2023,Querejeta2024} find little or no systematic difference.

\subsection{Clustering of star formation} \label{sec:cluster_sf}
\changeSummary{{This section is new and not contained in the former version of the paper. We expanded the paper by an analysis of the grouping of star formation and SN, as grouped star formation, that results in grouped stellar feedback, is essential for the evolution of a galaxy and assumptive heavily influenced by the Galactic potential.}}

\begin{figure*}
    \includegraphics[width=0.97\textwidth]{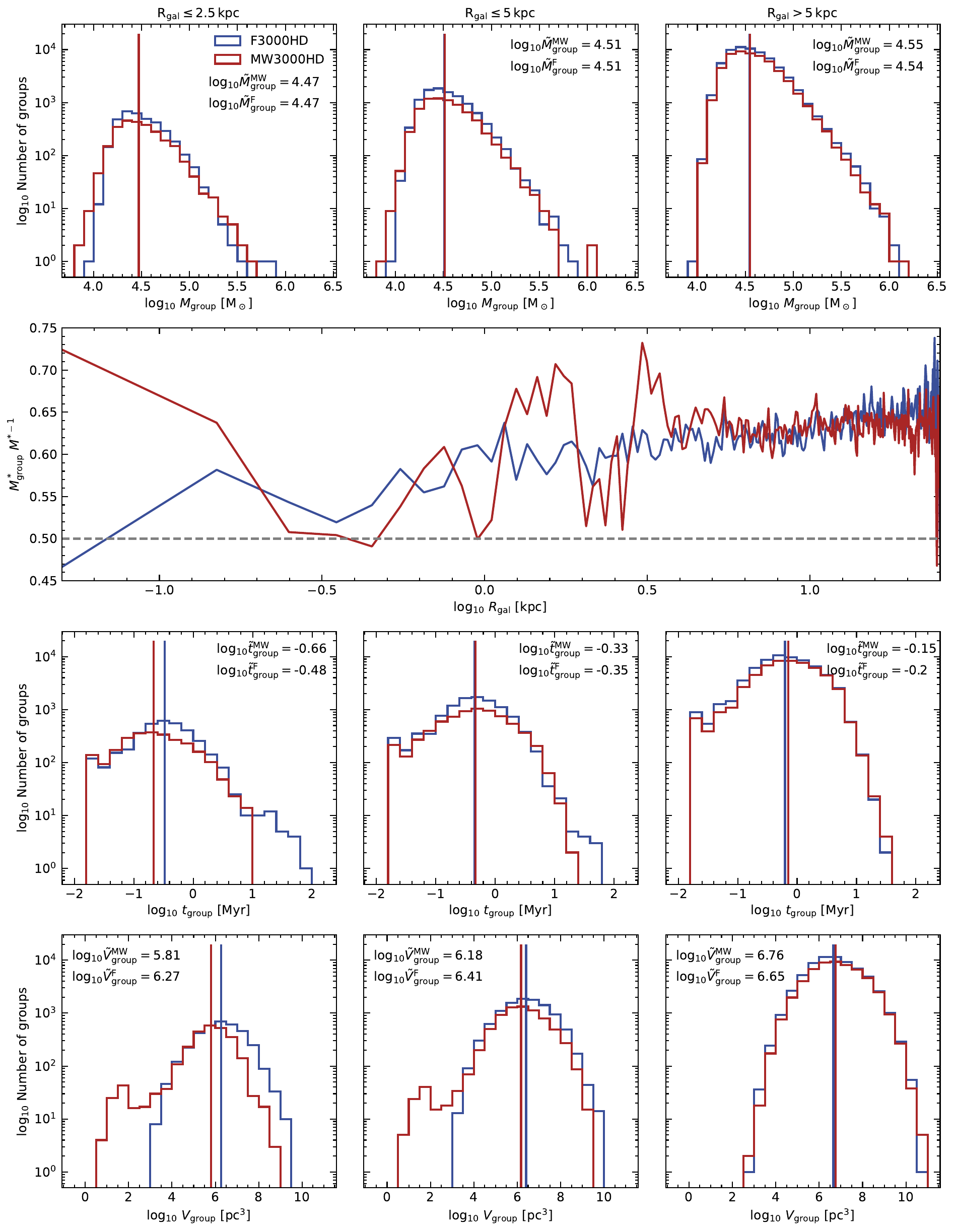}
    \caption{Grouped star formation: Mass of group (first row), mass fraction of stars born in groups (second row), activity time (third row) and extend of groups (fourth row) for F3000HD (blue) and MW3000HD (red) in Galactic center ($\leq2.5$~kpc, left column), inner region ($\leq5$~kpc, middle column) and disk ($>5$~kpc, right column). Vertical lines indicate median values, which are written out at the upper left or right.}
    \label{fig:clustering_starp}
\end{figure*}

In Section \ref{sec:rad_sf} and \ref{sec:az_sf} we found an increased concentration of SF activity towards the Galactic center and a more coherent azimuthal distribution of SF over time in MW3000HD compared to F3000HD. This raises the question whether in those regions of increased SF the star particles also form in a more clustered manner. If that is the case, we are also interested in the properties of those clusters. Also, a clustered SF also would have consequences for stellar feedback, which we study in section \ref{sec:cluster_sn}.

For this analysis we use the python clustering library \textsc{hdbscan} \citep{Compello2013}. \textsc{hdbscan}, together with other clustering algorithms, was tested on Gaia data to detect open clusters in five dimensional space of coordinates, proper motion and rotation speed by \citet{Hunt2021}. It was found to have highest sensitivity of the tested algorithms and in general to work best for Gaia data. \textsc{hdbscan} is a hierarchical clustering algorithm and is able to find clusters in regions of vastly different densities, utilizing a minimum spanning tree weighted by nearest-neighbour distances. Via single-linkage, it is then checked if the termination of a single connection would divide a cluster into two clusters, each with a number of members greater than a predefined number (\texttt{min\_cluster\_size}), or if the point would be merely falling out of the cluster. In the first case, the new clusters are kept as individual clusters, in the latter case the cluster keeps its structure. The result is a cluster hierarchy, from which the flat clustering is then derived by checks for stability of the clusters, again based on the weighted distances. \textsc{hdbscan} therefore does not utilize some predefined density-threshold or similar, but instead finds stable overdensities compared to the local background.
The algorithm is controlled by the single parameter \texttt{min\_cluster\_size} mentioned before, setting the lowest number of members a cluster can have. 
In our analysis, we set the parameter \texttt{min\_cluster\_size} to 5, i.e. we do not take into account any associations with fewer than five star particles. We are looking for groups in a four dimensional space of birth coordinates and birth time, so star particles identified to be in the same group have to be formed in close spatial vicinity and soon after each other. The found groups have very similar distributions in the number of group members across simulations and galaxy regions, ranging from the defined minimum number of 5 to several hundreds of star particles. The median number of group members is 10 and therefore well above the selected minimum number.

We are simply interested in the properties of regions where stars are not formed in isolation, where those regions are, what their size is and for how long they last. Therefore, we do not name the found structures `clusters' but simply `groups', as `cluster' is a widely used term in astronomical context. However, the associations found with \textsc{hdbscan} are not meant to resemble stellar clusters, as each star particle itself already represents several (massive) stars (see Section \ref{sec:methods:sf} and \ref{sec:methods:sfb}).

We present our findings for different regions of the simulated galaxies in Fig. \ref{fig:clustering_starp}. From left to right we select different radial ranges: $R_\mathrm{gal}\le2.5\,\mathrm{kpc}$ (left), $R_\mathrm{gal}\le5\,\mathrm{kpc}$ (middle), $R_\mathrm{gal}>5\,\mathrm{kpc}$ (right).
Because of the aforementioned similar distribution in number of group members, and because star particles in our simulations have little spread in mass, the mass distributions of groups are also very similar (see Fig. \ref{fig:clustering_starp}, first row) across simulations and Galactic regions.
They range from about $10^4$ to $10^6$~M$_\odot$, with a median at about $10^{4.5}$~Msun and a tail towards higher masses.

\begin{figure}
    \centering
    \includegraphics[width=\linewidth]{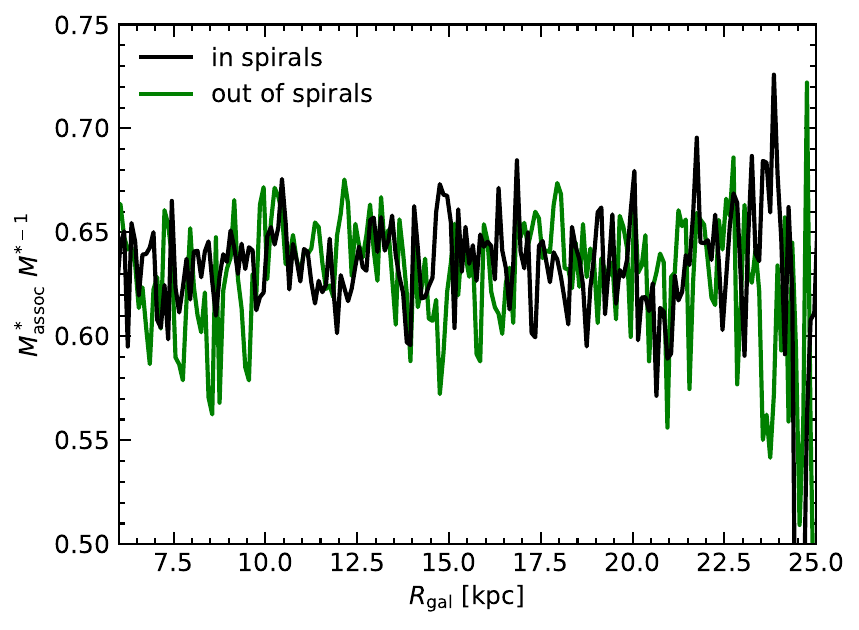}
    \caption{Fraction of stellar mass formed in associations within spirals arms (black) vs in interarm regions (green).}
    \label{fig:clustering_spirals}
\end{figure}

We also present the stellar mass fraction born in groups $M_\mathrm{group}^* / M^{*}$ (Fig. \ref{fig:clustering_starp}, second row). We find that in general more than 50~\% of the stellar mass is formed in groups, for the innermost 5~kpc ranging between $\sim$50~\% to $\sim$70~\%, and being relatively stable at around 65~\% in the disk region (with large statistical fluctuations at the Galactic outskirts). In the inner 5~kpc of F3000HD we can see an increasing grouping with radius, but no clear trend for MW3000HD, and also no enhanced grouping in SF is present in Galactic spirals, as one might expect (see Fig. \ref{fig:clustering_spirals}). We therefore conclude clustering to be an ubiquitous phenomenon, present at a comparable fraction in all regions of the Galaxy.

In the third row of Fig. \ref{fig:clustering_starp}, we present the activity time $t_\mathrm{group}$ of the group, which is calculated as the difference between the birth time of the first and last star particle of the group. It therefore represents the time over which a star forming region actively forms star particles. Star forming regions with activity times lower than the global time step we group together in the lowest bin depicted. In the activity times, differences between the Galactic regions are present: the median activity time increases with galactocentric distance, being larger in the Galactic disk than in the central region. This means that star forming regions consisting of multiple star forming cells persist for longer in the Galactic disk than in the center. Finally, we also find a difference between Galactic models in the Galactic center, with groups in the innermost 2.5~kpc (i.e. in the inner bar region) of MW3000HD having an about $0.2$~dex lower activity time than in F3000HD, whose distribution shows a tail towards long activity times. This probably is caused by stronger shear, which we know to influence SFRs \citep[e.g.][]{Colling2018} and turbulence because of the bar potential in MW3000HD.

Another result of higher shear is the prevalence of smaller groups in the center of MW3000HD than in F3000HD (see Fig. \ref{fig:clustering_starp}, last row). `Smaller` here means a lower occupied cuboid volume $V_\mathrm{group}=\left(\max(x)-\min(x)\right)(\max(y)-\min(y))(\max(z)-\min(z))$ (compared to the large dynamical range of 10 orders of magnitude, the actual measure of the occupied volume, be it cuboid, spherical or a convex hull, is negligible). The median extent in MW3000HD within the innermost 2.5~kpc is about $0.4$~dex lower than in F3000HD, with a noticeable tail towards low volumes. In the Galactic disk, no such difference between simulations is present, and median group extends are up to $1$~dex more extended than in the Galactic center.

In summary, the star clustering properties are heavily affected by the external potential in the central region, where the groups of stars are more compact and form faster in the MW than the flat potential. On the other hand, the groups of stars formed outside the central regions have the same statistical properties which thus does not depend on the external potential.

\subsection{Clustering of supernova feedback} \label{sec:cluster_sn}
\changeSummary{{This section as well is new and not included in the previous version of the paper.}}

\begin{figure*}
    \centering
    \includegraphics[width=\textwidth]{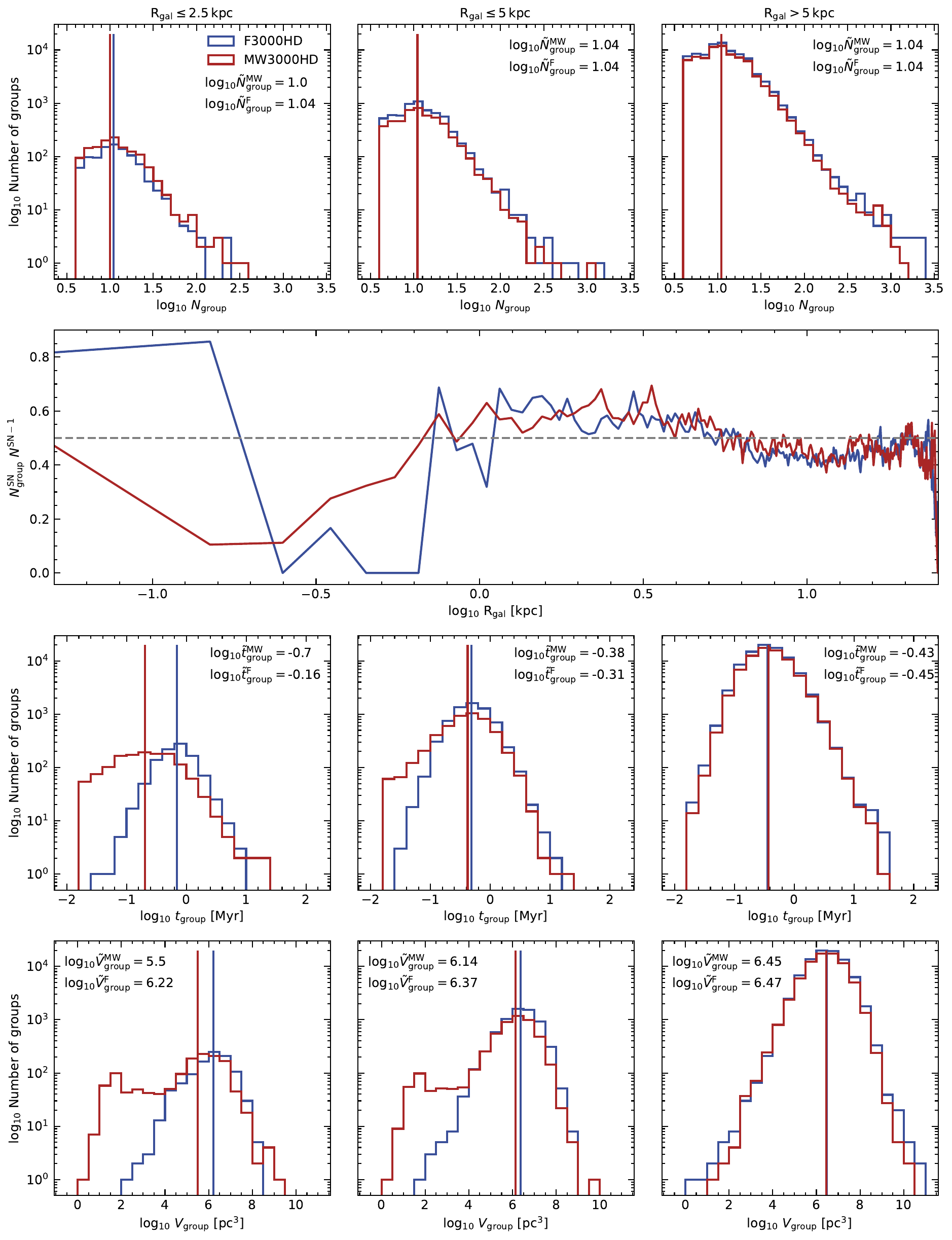}
    \caption{Grouped supernovae: Number of association members (top row), fraction of SN exploding within groups (second row), activity time (third row) and extend of groups (fourth row) for F3000HD and MW3000HD for t=2450-2550~Myr. Vertical lines indicate median values.}
    \label{fig:clustering_sn}
\end{figure*}

As soon as star particles formed from the gas, they are decoupled from the hydrodynamic flow and are subject just to gravitational forces. This increases the impact the gravitational potential has on their motion, making stellar feedback, as it is just tied to the dynamics of star particles, an interesting probe of the Galactic potential. Moreover, SNe exploding in groups have a substantial impact on their surrounding environment, as they can form superbubbles, which is not possible for individual SNe.

Because of the high number of supernova explosions (each star particle can hold several massive stars that explode as SNe), we can only perform the analysis for a limited time, which we choose to be 2450-2550~Myr. While  we find strictly more than 50~\% of stellar mass being formed in groups in the Galactic center, the number fraction of SN exploding in groups in the center ($5\approx10^{0.7}$~kpc) is very volatile and ranges from 0 to about 60~\% (see Fig. \ref{fig:clustering_sn}, second row). Further out, the fraction falls below 50~\% (where it is about 65~\% for stars formed). This does not differ between MW3000HD and F3000HD. This suggests that groups are ripped apart in between the formation of stars and the explosion times of SNe. In the innermost 1~kpc, SNe are grouped to a lower degree in space and time in MW3000HD than in F3000HD (for this small radius we were able to check grouping for a longer timespan, giving a more clear result, but we show data of the selected timespan here for consistency).
This shows that a barred potential is necessary to get the right clustering of stellar feedback in the Galactic center.

As for star particles, groups of SNe in the center of MW3000HD experience shorter activity times than in F3000HD ($\sim 0.54$~dex), whereas activity times do not differ in the Galactic disk $>5$~kpc (third row; again we collect all clusters with an activity time below the global time step of the simulation in the lowest bin). In the disk the median activity time of SN regions is also $0.2-0.3$~dex lower than for star particle groups.

SN groups in the Galactic center have about $0.7$~dex smaller mean extent in MW3000HD than in F3000HD, with the size distribution of MW3000HD showing a similar, but even stronger bimodal pattern than observed for the star particles. In the disk, no such difference is present. Extends of SN groups in general are slightly lower than those of star particle groups.

\begin{figure*}
    \centering
    \includegraphics[width=\textwidth]{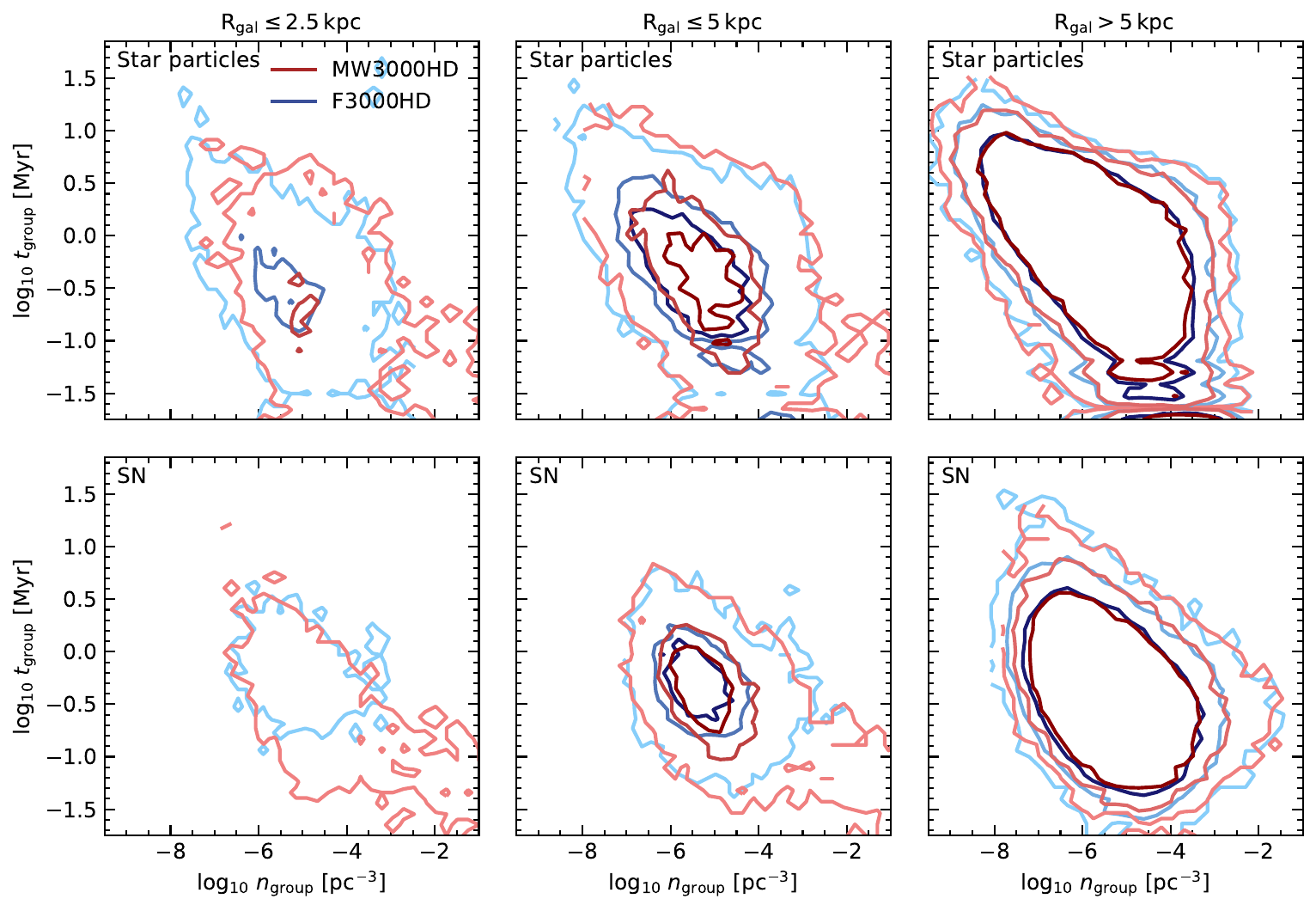}
    \caption{Activity times of groups of star particles (upper row) and SN (lower row) vs their density for MW3000HD and F3000HD. Contour lines indicate levels of 1, 25 and 50 star particles/SN.}
    \label{fig:cluster-dens-lifetime}
\end{figure*}

To shed some light on the connection between the activity time of a group and its other properties, we present the activity time plotted vs the number density $n_\mathrm{group}=N_\mathrm{group}V_\mathrm{group}^{-1}$ for star particles and SN in Fig. \ref{fig:cluster-dens-lifetime}. While we find a positive correlation between activity time and number of group members (i.e. groups with more members need longer to form and have a larger timespan in which SN explode) and activity time and group volume (more extended groups need longer to form and have a longer timespan in which SN explode), there is actually a negative correlation between activity time and number density in both star particles and SN. This means that groups with more members per volume form faster. This is instructive, as those regions in the simulation resemble star forming clouds composed of multiple cells. The denser (and smaller) this cloud is, the higher is the probability for its cells to be converted into a star particle. Groups that form within a short timespan also have their SN explode within a shorter time, as the time offset due to delay between the formation of star particles is low. The delay between SN then is mainly caused by different masses and therefore lifetimes of the individual stars enclosed in the star particles.

\section{Caveats} \label{sec:caveats}
\changeSummary{{Section "Missing Physics" was moved here.}}

Some of the presented results reveal shortcomming of our simulations, which, however, can never be fully avoided in simplified models of complex systems.

The global SFR, for example, shows a continuous decline over the simulated time. In order to study the prevailing processes in our simulations precisely, we do not wish to account for gas accretion onto the Galactic disk, and therefore model our Galaxies without a circum-galactic medium. The only gas accretion onto the ISM is therefore previously expelled gas from the Galactic disk. The lack of additional supply of fresh gas, however, results in a progressive depletion of gas in the simulation, as gas gets converted into star particles. This, in part, is a desired result, but also leads to a limited time over which the results are directly comparable to the Milky Way. The lack of gas at late times also results in a SFR below the typical MW value of 1~M$_\odot$~yr$^{-1}$.

The isolated-galaxy approach also results in a lack of satellites for our Milky Way analogues. The observed Milky Way has multiple smaller satellite galaxies, with the Large and Small Magellanic Clouds being the largest ones, which influence the Galactic dynamics (see e.g.\ \citealt{LMC-MW}) as well as the evolution history (e.g.\ \citealt{MW-merger-sim} and \citealt{MW-merger-obs}). In today's Milky Way observations, past mergers are still present as stellar streams or globular clusters. This lack of companions (as well as a low age) might be the reason for the thinness of the stellar and gaseous disk in our simulations.

In addition, we also ignore pre-supernova feedback such as photoionization and stellar winds in our simulations. This might impact the ISM structure in terms of increasing the amount of ionized gas. However, we plan to include them in future work. 

In this paper we focus on the hydrodynamic simulations omitting magnetic fields and also cosmic rays, both processes are known to have an impact on the star formation and the dynamics of the ISM. Simulations including magnetic fields and cosmic rays will be discussed in companion papers (Kjellgren et al., subm.; Girichidis et al., in prep.).

\section{Conclusions}
\label{sec:conclusions}
\changeSummary{Summary was remodeled to fit the new structure of the paper.}

In this work we introduce the Rhea simulation suite, a set of comparable simulations of Milky Way-like galaxies with varying external potential. We study the influence of a detailed external potential tuned to recreate observations of morphology and dynamics of the Milky Way. We focus on studying the influence of the potential on star formation and stellar feedback.

Our main finding is that while the precise shape of the potential does not change the overall averaged SFR properties, it changes where star formation takes place. Indeed, the MW external potential favors star formation in long-lived spiral arms and ensures an continuous fueling of the central zone. Furthermore, the external MW potential influence the properties of star forming regions in the Galactic center, producing compact groups that forms faster than with a flat potential.

Our detailed findings are:
\begin{itemize}
    \item Both simulations follow a temporal evolution that is close to steady state conditions, i.e. the rotation speeds are very close to the analytical ones defined via the external potential, and the continuous formation and evolution of spiral arms. The star formation rate decreases with the continuous depletion of gas without strong temporal fluctuations. 
    \item Apart from the central region around the Galactic bar, the overall morphology of the galaxy is hardly influenced by the imposed potential: both simulations show similar spiral structures and feedback driven bubbles. 
    \item In the MW potential, the vertical spread of both stars and gas in the bar region is higher than in the flat potential. This is because the external MW potential is weaker than the external flat potential in the central 5~kpc of the galaxy.
    \item The potential has no noticeable influence on the global SFR, but the MW potential, in particular the bar, prevents the Galactic center from quenching over time. In this potential, SF is strongly centered and strongest in the innermost 1.5~kpc.
    \item In the MW potential, the density of gas, stars and hence star formation tend to increase in the potential wells of the spiral arms and bar. Approximately 60~\% of stars outside the bar region form within spiral arms. Such a pattern is not observable in the flat external potential, no long-lived star forming spiral patterns form.
    \item Most star particles form close to other star particles in groups in both simulations. Shear forces of the bar potential lower the lifetime and size of those associations compared to a non-barred potential.
    \item Most SNe explode in isolation. The activity times and sizes of associations of SNe are lowered in the bar region of the MW potential compared to the flat potential.
\end{itemize}

As a conclusion, it is therefore sufficient to use a simplified axisymmetric potential to reproduce the overall star formation properties of the  Milky-Way. However, if one is interested in the distribution of star formation within the galaxy, 
or if one is interested in the Galactic center in particular, then the use of a barred potential is indispensable.

\begin{acknowledgements}
This research was funded by the European Research Council via the ERC Synergy Grant ``ECOGAL'' (project ID 855130).
The team in Heidelberg acknowledges support from the German Excellence Strategy via the Heidelberg Cluster of Excellence (EXC 2181 - 390900948) ``STRUCTURES'', and from the German Ministry for Economic Affairs and Climate Action in project ``MAINN'' (funding ID 50OO2206).
The authors gratefully acknowledge the scientific support and HPC resources provided by the Erlangen National High Performance Computing Center (NHR@FAU) of the Friedrich-Alexander-Universität Erlangen-Nürnberg (FAU) under the NHR project a104bc. NHR funding is provided by federal and Bavarian state authorities. NHR@FAU hardware is partially funded by the German Research Foundation (DFG) – 440719683.
They also thank for computing resources provided by the Ministry of Science, Research and the Arts (MWK) of the State of Baden-W\"{u}rttemberg through bwHPC and the German Science Foundation (DFG) through grants INST 35/1134-1 FUGG and 35/1597-1 FUGG, and for data storage at SDS@hd funded through grants INST 35/1314-1 FUGG and INST 35/1503-1 FUGG. NB acknowledges support from the ANR BRIDGES grant (ANR-23-CE31-0005). MCS acknowledges financial support from the European Research Council under the ERC Starting Grant ``GalFlow'' (grant 101116226) and from Fondazione Cariplo under the grant ERC attrattivit\`{a} n. 2023-3014. 
JG, KK and GH are fellows of the International Max Planck Research School for Astronomy and Cosmic Physics at the University of Heidelberg (IMPRS-HD). RSK is grateful for support from the Harvard Radcliffe Institute for Advanced Studies and Harvard-Smithsonian Center for Astrophysics for their hospitality and support during his sabbatical. JG wants to thank Matthew Smith, Rüdiger Pakmor and David Whitworth for helpful and insightful discussions. For this work, we used astropy \citep{astropy1, astropy2, astropy3}, yt \citep{yt}, numpy \citep{numpy}, matplotlib, scipy \citep{scipy}, AGAMA \citep{agama} and Arepo \citep{arepo}.
\end{acknowledgements}

\bibliography{references, global}{}
\bibliographystyle{aa}

\clearpage
\appendix

\section{Outcome of phase I}
\changeSummary{Removal of MHD and some rework.}

Before starting the simulation we aim to analyze, our simulated galaxies undergo an initial development phase that we call phase I. During this phase, the galaxies develop from a smooth gaseous density distribution to a turbulent and structured distribution which we start our final simulation from.

\begin{figure*}[t!]
\centering
\includegraphics[width=\textwidth]{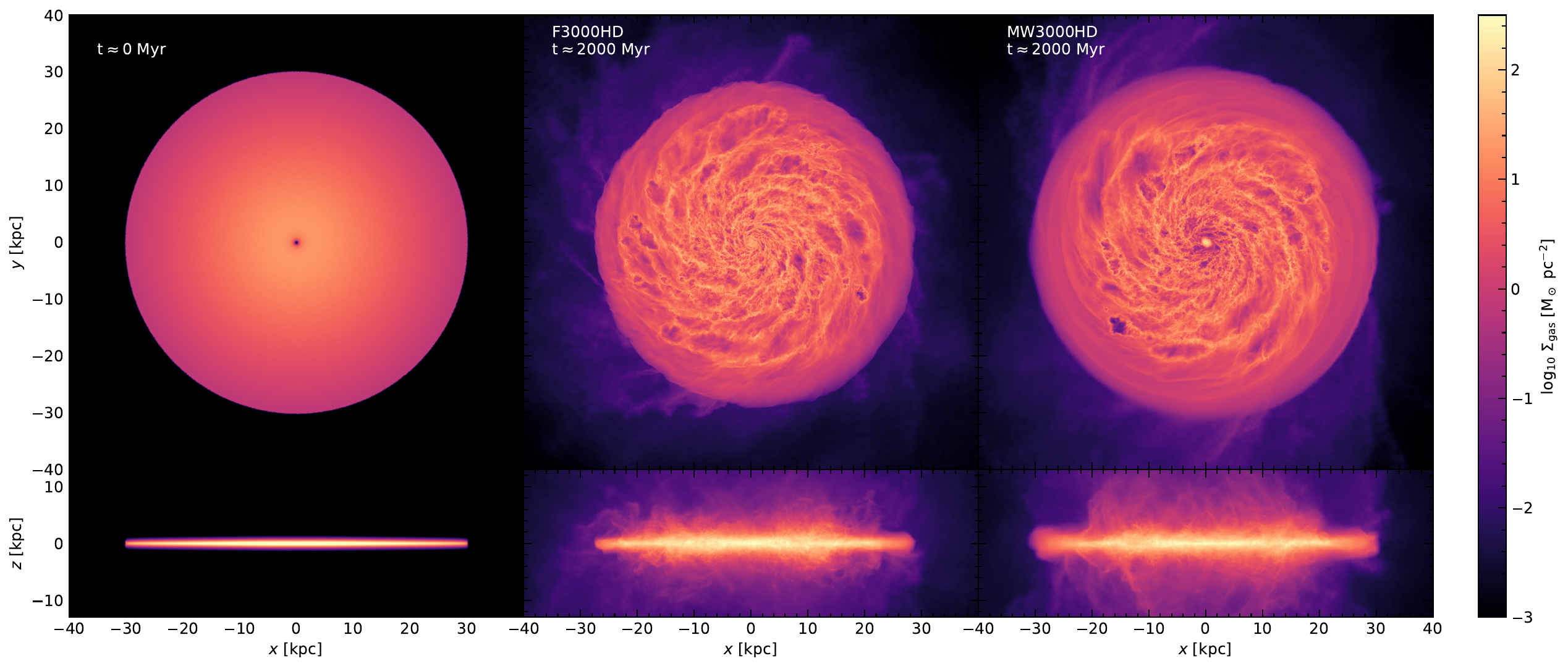}
\caption{Projected gas surface density at the end of phase I for all simulations, as well as at $t=0$~Myr (initial conditions).}
\label{fig:preproc:flat}
\end{figure*}

We show these start and end points of phase I in Fig.\ \ref{fig:preproc:flat}. The panel in the upper left depicts the initial smooth density distribution following Equation \ref{equ:density_profile} in a projection along the y and z axis. The following panels show the developed state at the end of phase I of our simulation runs. The general behavior also observed in Fig.\ \ref{fig:morphology_plot} is already prevalent at this time.

Due to numerical limitations in the mesh refinement, during that phase we lose some SN which explode in regions with strongly diverging gas motion. This applies to 1.8~\% of all SN in F3000HD and 6.3~\% in MW3000HD. During phase II, which is the phase we analyze, this fraction lowers to about 0.6~\% for both simulations.

\section{Resolution study}
\changeSummary{Former Fig. 2 added here.}

As a test of convergence we conducted simulations of both potentials also with a mass resolution 1000 M$_\odot$. Overall we see very good agreement with our fiducial runs F3000HD and MW3000HD. 

\begin{figure*}
    \centering
    \includegraphics[width=0.97\columnwidth]{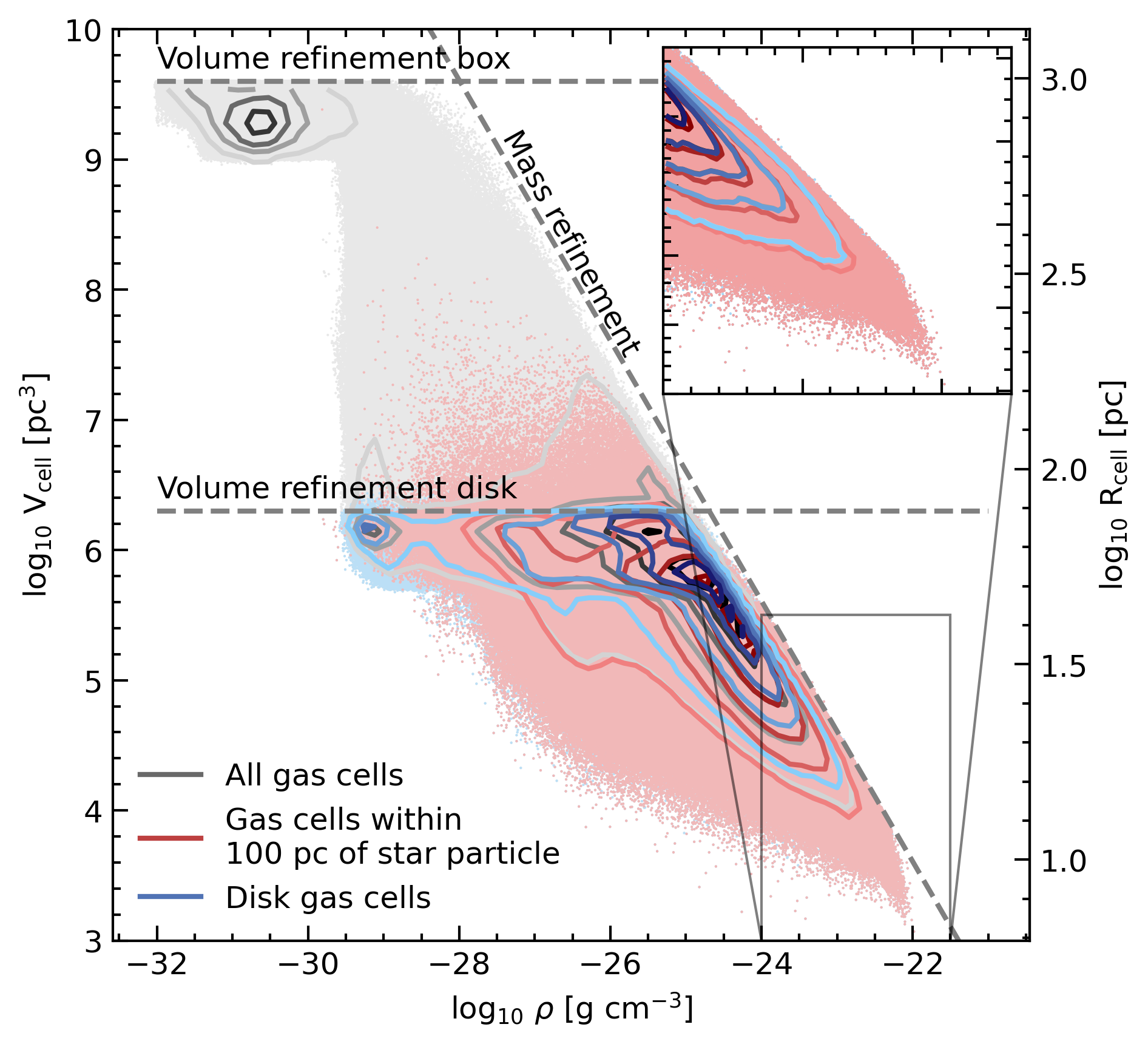}
    \includegraphics[width=\columnwidth]{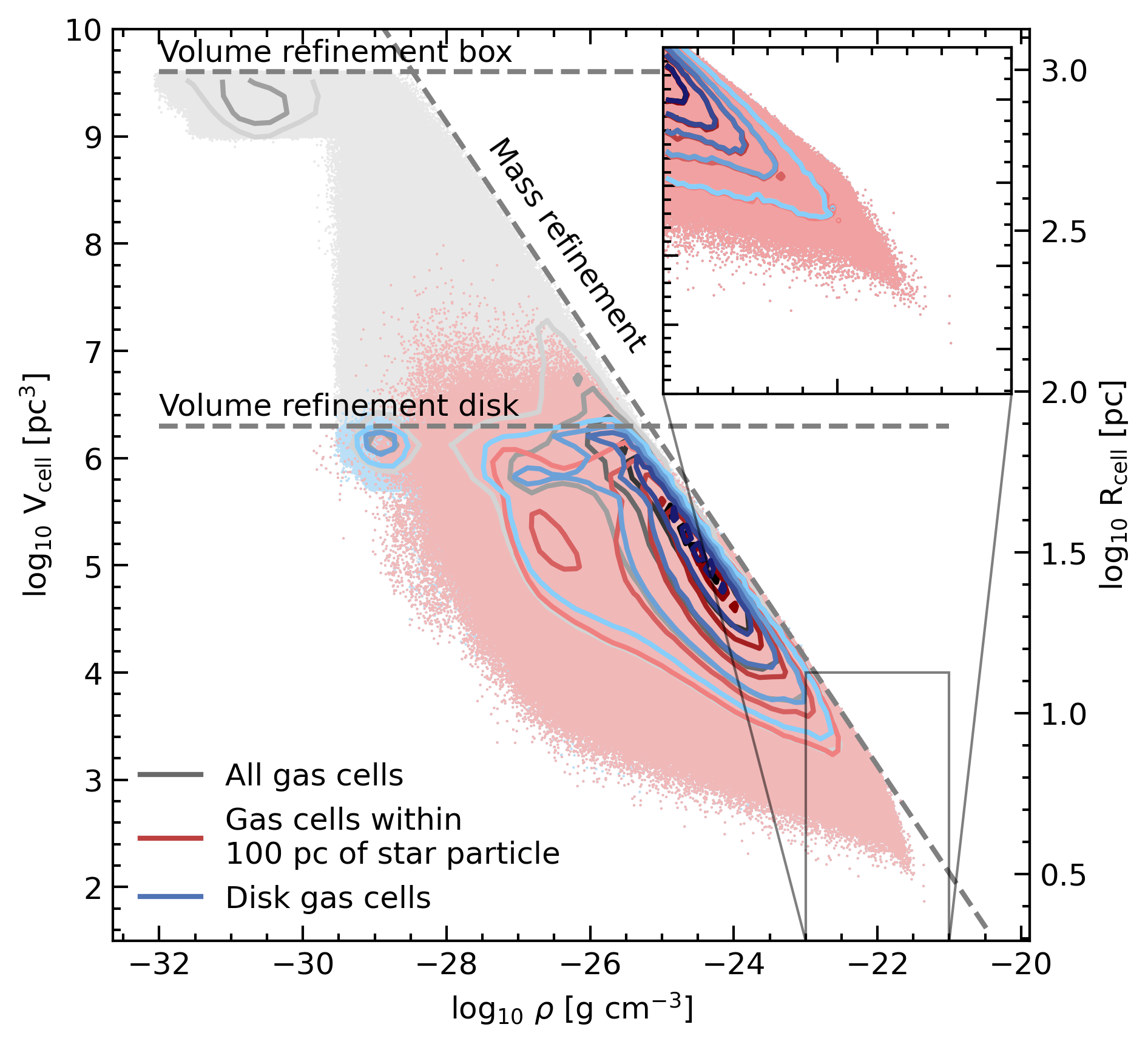}
    \caption{Volume-density distribution of cells in F3000HD (left) and F1000HD (right) at about 2500~Myr. Contour levels enclose 10\%, 30\%, 50\%, 70\% and 90\% of cells.}
    \label{fig:density-volume_with_highres}
\end{figure*}

In Fig.\ \ref{fig:density-volume_with_highres} we again show the correlation between cell size and cell density, for F1000HD and F3000HD at about 2500~Myr in phase II. It is apparent that in the higher resolution simulation the cells get to lower sizes and correspondingly higher densities due to the lower target mass (up to $\sim10^{-21.4}$~g~cm$^{-3}$ compared to $\sim10^{-22}$~g~cm$^{-3}$)

\begin{figure*}
    \centering
    \includegraphics[width=\textwidth]{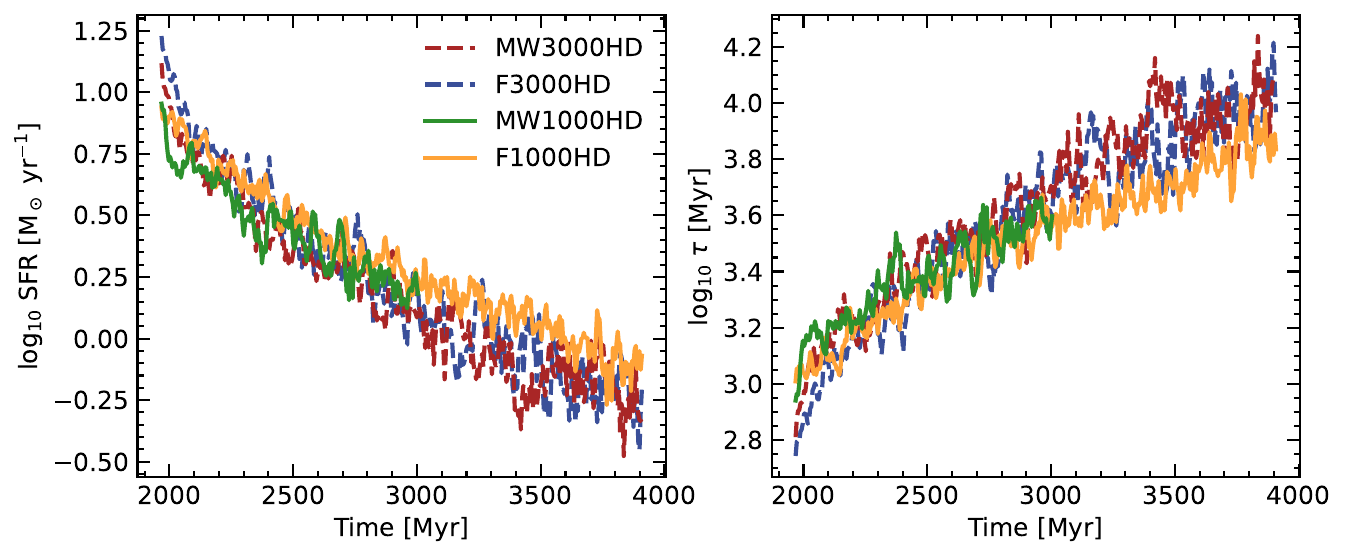}
    \caption{SFR history (left) and depletion time (right) for F3000HD (blue), MW3000HD (red), F1000HD (orange) and MW1000HD (green). We limit the measurement to $\pm1$~kpc above and below the Galactic plane. We find good agreement in SFR between simulations with 3000~M$_\odot$ and 1000~M$_\odot$ target mass.}
    \label{fig:sfr_with_highres}
\end{figure*}

In Fig.\ \ref{fig:sfr_with_highres} we present the SFR and depletion time $\tau$ of the fiducial simulations F3000HD and MW3000HD (same as Fig.\ \ref{fig:SFR}), overlaid with the curves of F1000HD (orange) and MW1000HD (green). Since running with 1000~M$_\odot$ resolution is more computationally costly, MW1000HD did just run up to about 3000~Myr, i.e. 1000~Myr into phase II. We see excellent agreement between the SFRs produced by F3000HD and F1000HD, as well as between MW3000HD and MW1000HD within the first 1000~Myr of phase II. Correspondingly, the depletion times of the simulations agree as well. At later stages of the simulations, SFR tends to be slightly higher at higher resolution, because of the increased ability of gas to collapse to high density. However, this effect is mild.

\begin{figure*}
    \centering
    \includegraphics[width=\textwidth]{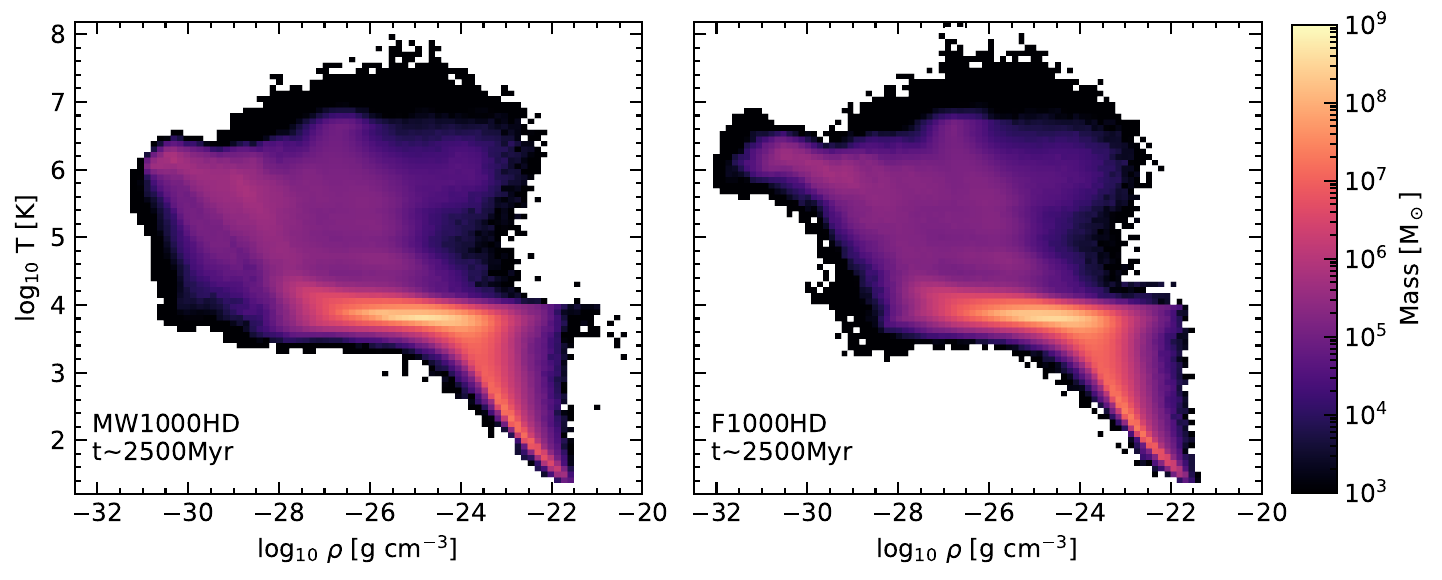}
    \caption{Temperature-density plot for F1000HD and MW1000HD at the fiducial analysis time of 2500~Myr. We find good agreement with Fig.\ \ref{fig:phaseplot}.}
    \label{fig:phaseplot_with_highres}
\end{figure*}

The distribution of gas in temperature and density (Fig.\ \ref{fig:phaseplot_with_highres}, same as Fig.\ \ref{fig:phaseplot}, but for MW1000HD and F1000HD) also shows no noticeable difference between 3000~M$_\odot$ and 1000~M$_\odot$ target mass in both potentials. Gas reaches slightly higher densities (as already described) and temperatures with lower target mass, but overall we see good convergence.

\section{Additional figures}
\begin{figure*}
    \includegraphics[width=0.97\linewidth]{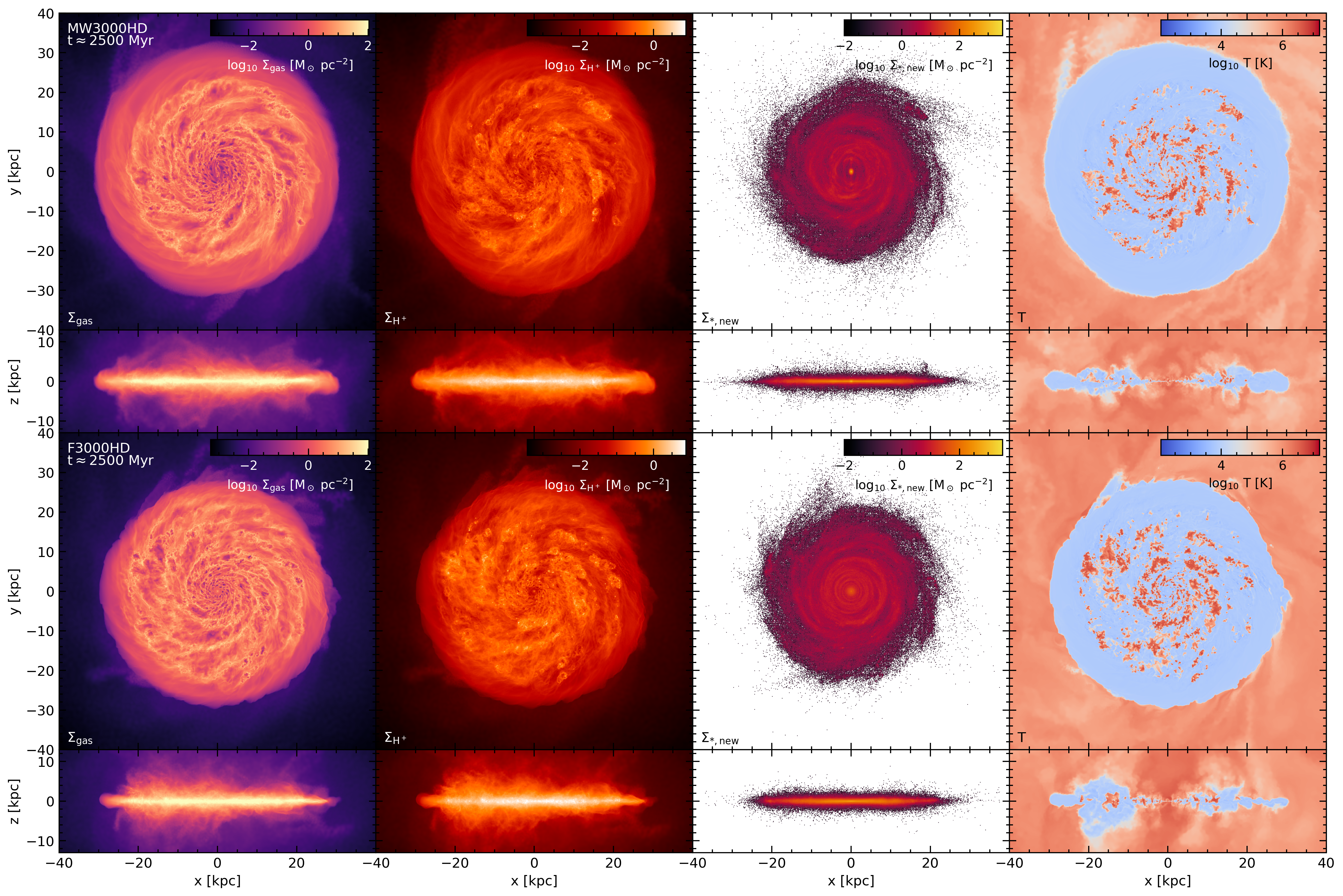}
    \caption{Same as Fig. \ref{fig:morphology_plot_center}, but zoomed out to see the whole Galactic disk and circum-galactic medium.}
    \label{fig:morphology_plot}
\end{figure*}
\end{document}